\documentclass[useAMS,usenatbib,usegraphicx]{mn2e}

\title[Formation histories and evolution of E+As]{Star formation histories and evolution of 35 brightest E+A galaxies from SDSS DR5}
\author[W. Du et al.]{W. Du,$^{1,2,3}$\thanks{Email: wdu@nao.cas.cn} A. L. Luo,$^{1,2}$ Ph. Prugniel,$^4$ Y. C. Liang,$^{1,2}$ Y. H. Zhao$^{1,2}$\\
$^{1}$National Astronomical Observatories, Chinese Academy of Sciences, Beijing, 100012, China\\
$^{2}$Key Laboratory of Optical Astronomy, National Astronomical Observatories, Chinese Academy of Sciences, Beijing, 100012, China\\
$^{3}$Graduate University of Chinese Academy of Sciences, Beijing, 100049, China\\
$^{4}$Universit ́ Lyon 1, Villeurbanne, F-69622, France; CRAL, Observatoire de Lyon, St. Genis Laval, F-69561, France ; CNRS, UMR 5574
}

\begin{document}

\date{}

\pagerange{\pageref{firstpage}--\pageref{lastpage}} \pubyear{}

\maketitle

\label{firstpage}
\begin{abstract}
We pick out the 35 brightest galaxies from Goto's E+A galaxies catalogue which are selected from the Sloan Digital Sky Survey Data Release 5. As E+As have experienced starburst recently and quenched it abruptly, they have been considered as post-starburst galaxies. The spectra of E+As are prominently characterized by the strong Balmer absorption lines but little [O{\sc{ii}}] or H${\rmn{\alpha}}$ emission lines. In this work we study the stellar populations of the sample galaxies by fitting their spectra using ULySS, which is a robust full spectrum fitting method. We fit each of the sample with 1-population (a single stellar population-a SSP) and 3-population (3 SSPs) models, separately. By 1-population fits, we obtain SSP-equivalent ages and metallicities which correspond to the `luminosity-weighted' averages. By 3-population fits, we divide components into three groups in age (old stellar population-OSP, intermediate-age stellar population-ISP, and young stellar population-YSP), and then get the optimal age, metallicity and population fractions in both mass and light for OSP, ISP and YSP. During the fits, both Pegase.HR/Elodie3.1 and Vazdekis/Miles are used as two independent population models. The two models result in generally consistent conclusions as follows: for all the sample galaxies, YSPs ($\le$ 1Gyr) make important contributions to the light. However, the dominant contributors to mass are OSPs. We also reconstruct the smoothing star formation histories (SFHs) by giving star formation rate (SFR) versus evolutionary age. All the sample galaxies have low SFRs in the intermediate evolutionary stage. Eleven of the thirty-five E+As have high SFRs in the early evolutionary stage. However, another 11 have SFRs that are high during the late evolutionary stage. This might be due to the recently happened but abruptly truncated starburst in such galaxies. In addition, we fit the E+A sample and 34 randomly selected elliptical galaxies with 2-population (2 SSPs) model, which could divide the galaxy components into two groups in age (old component and young component). We obtain the equivalent age of old components for each of the E+A sample and elliptical galaxies. By comparison, the old components of E+As are statistically much younger than those of ellipticals. From the standpoint of the stellar population age, this probably provides an evidence for the proposed evolutionary link from E+As to early-types (E/S0s).

\end{abstract}


\section{Introduction}

There exists such a class of galaxy as has optical spectrum characterized by strong Balmer absorption lines, but no or insignificant emission in [O{\sc{ii}}] or H${\rmn{\alpha}}$. These galaxies are called `E+As' or `K+As' , for their spectra look like a superposition of passive continuum spectra of eliptical galaxies or an ensemble of K-type stars (Mg$_{\rmn{5175}}$, Fe$_{\rmn{5270}}$ and Ca$_{\rmn{3934,3468}}$ absorption lines but no emission lines) and of A-type stars (strong Balmer absorption lines) \citep{b33}. Therefore, the so-called E+A or K+A galaxy is only a classification of spectrum but not a classification of morphology. The existing of strong Balmer absorption lines in the spectra indicates the recently experienced starburst or the presence of young stars in these galaxies. However, no detection of emission line in [O{\sc{ii}}] rules out the ongoing star formation in these galaxies. So E+As are considered as post-starburst galaxies. 

It is proposed that E+As are good candidates in the middle stage of the transformation from late-types (star-forming, gas-rich, disk-dominated galaxies) to early-types (quiescent, gas-poor, spheroid-dominated galaxies). They locate at such a crucial evolutionary stage that it is essential to study them for a better understanding of the galaxy evolution and for exploration of the origin of the red sequence of galaxies. So far, studies on E+As and the evolutionary link with early-type galaxies have been presented in several ways.  

\citet{b13,b20,b12} have presented 21 E+As from high-resolution HST ACS and WFPC2 images. Based on surface photometry and 2-dimensional model fit, they studied the morphologies, color profiles, new star clusters formed during the starburst, the location on the fundamental plane (FP) of elliptical galaxies, and the evolution of the FP zero point. By disturbed morphologies and locations in the field, they suggested that galaxy-galaxy mergers and interactions are one mechanism which triggers off the E+A phase. The discovery of LINER spectra in the cores of some E+As in the follow-up spectroscopy gave confidence in the proposed evolutionary link (E+As $\rightarrow$ E/S0s).

In \citet{b14}, they have presented 10 relatively bright (b$_{\rmn{J}}$ $\sim$ 18.4) and nearby (z $<$ 0.2) E+A galaxies from 2dFGRS imaging and spectroscopic observations. They made combined photometric and spectroscopic study for the 10 E+As. In photometry, the structures, morphologies and spatial distributions of colours have been investigated. Morphologically the sample galaxies have taken on a look of tidal features, which indicated a high rate of interactions. In addition, the radial surface brightness profiles of the sample galaxies have enhanced the convincingness to the proposed evolutionary link (E+As $\rightarrow$ E/S0s). In spectroscopy, stellar populations have been constructed by fitting spectra with template. Young stellar populations they have obtained for all the sample galaxies showed significant rotation, and this phenomenon commonly existed among overall populations of early-types.

\citet{b29} compared the spectra of E+A galaxies with the template of E and S0 spectra, and this comparison indicated that early-type galaxies are successors of E+A galaxies. They used the galaxy evolution code to trace the evolution of the Lick indices and colors of the E+A galaxies. In terms of internal structures of galaxies, \citet{b30} have studied the radial and 2D colour properties of 22 E+As from Sloan Digital Sky Survey Data Release 2 \citep{b40} and suggested that E+A galaxies are post-starburst galaxies which have undergone a centralized starburst arising from interactions and mergers. \citet{b31} compared the fundamental planes (FP) of E+A galaxies with FP of UV-excess early-type galaxies. Their work suggested that E+A galaxies might undergo the transition from `blue cloud' to `red sequence', and eventually migrate to the red sequence early-type galaxies .

More studies will rely on E+As, so a large number of E+A galaxies are needed from various telecopes and surveys. So far, there are already several E+A catalogues which are separately selected from Sloan Digital Sky Survey Data Release 5 \citep{b41} by \citet{b1}, the 2dF Galaxy Redshift Survey (2dFGRS) by \citet{b25}, the Las Campanas Redshift Survey (LCRS) by \citet{b21} and so on. \citet{b25} has selected sample of low-redshift (z $\sim$ 0.1) E+A galaxies from the 221,000 galaxy spectra in 2dFGRS. They adopted two different selection techniques, based on three Balmer absorption lines (H${\rmn{\delta}}$, H${\rmn{\gamma}}$, H${\rmn{\beta}}$) or solely the H${\rmn{\delta}}$ line. These two methods yield 56 and 243 E+As, respectively. \citet{b21} identify 21 low-redshift (z $\sim$ 0.05 - 0.13) E+As from the 11,113 galaxies in the LCRS by using criteria that the average of equivalent widths of H${\rmn{\beta,\gamma,\delta}}$ is greater than 5.5 $\rmn{\AA}$ and the equivalent width of [O{\sc{ii}}] emission line is less than 2.5$\rmn{\AA}$. Our sample is taken from Goto's E+A catalogue, and it will be described in next section. 

As stellar populations of galaxies hold a wealth of information about the formation history and evolution, we shall study stellar populations of the E+A sample. In addition, we expect to explore new pieces of evidence in the aspect of stellar populations for the proposed evolutionary scenario of E+As. At present, stellar populations can be widely constructed by the stellar population synthesis technique which is based on fitting photometric indices and colours or spectroscopic features with stellar population models. In this work, we choose 35 brightest E+A galaxies from Goto's catalogue \citep{b1}, and then use ULySS, which is a stellar population synthesis code \citep{b11}, to study the stellar populations and star formation histories. Comparing with the previous works on E+As , our work is based on spectroscopy instead of photometry since spectra contain much wealthier information than images. Our sample will be larger in number, and taken from SDSS spectroscopic catalogue, which has acknowledged higher quality spectra. 

In this paper, we describe our E+A sample in $\S$ 2. We model the stellar populations for the sample and make conclusions on dominance of stellar populations in $\S$ 3. In $\S$ 4, we verify the reliability and robustness of our solutions. We present star formation rates (SFRs) to reconstruct the smoothing star formation histories (SFH) for the sample in $\S$ 5. Additionally, in $\S$ 6 we provide a piece of evidence for the proposed evolutionary scenario of E+As. We summarize the work in $\S$ 7.

\section[]{Description of the E+A sample}
\citet{b1} has established a catalogue of 564 E+A galaxies identified from all galaxy spectra of the SDSS DR5 \citep{b41}. These E+As are selected by the following criteria: 

(1)spectroscopic signal-to-noise (S/N) ratio $>$ 10 per pixel (in the continuum of the r-band wavelength range)

(2)EW(H${\rmn{\delta}}$)$>$5.0$\rmn{\AA}$, EW(H${\rmn{\alpha}}$)$>-$3.0$\rmn{\AA}$ and EW([O{\sc{ii}}])$>-$2.5$\rmn{\AA}$ (EW is short for `equivalent width' and EWs of these three lines are measured by \citet{b42}. Here absorption line has a positive sign.) 

(3)galaxies with redshift between 0.35 and 0.37 should be excluded due to the sky feature at 5577$\rmn{\AA}$ after (1) and (2). 

By these criteria, the redshifts, absolute magnitudes and diameters of the whole E+A galaxies in this catalogue of \citet{b1} have distribution histograms as shown in Figure 1 (the dashed lines).

We picked the 35 brightest E+A galaxies as our sample from Goto's catalogue. We chose this sample to ensure that the objects are sufficiently bright for further spectroscopic study in our work. These 35 galaxies are nearby and the redshifts are all within 0.14 (z$<$0.14). They have Petrosian magnitudes between 14 and 16 in r-band, span a redshift range from 0.0381 to 0.1322 (Figure 1, the solid line in the left histogram), and have high spectroscopic S/N ratio in the continuum of the r-band wavelength range (S/N $>$ 30). In Table 1, we list the major properties of the sample.

In Figure 1, we present the distribution histograms of the full 564 E+As in \citet{b1} and the selected sub-sample in terms of redshift, absolute magnitude in mag, diameter in kpc, and diameter in arcsec. For easy comparisons, we plot the histograms of the full 564 (dashed lines) galaxies and 35 sample galaxies (solid lines) in the same picture in distinctive line styles. Particularly, we have transformed distributions of the 35 sample galaxies to the equivalent distributions if the number of 35 is enlarged to 564 . This transformation is just for a more direct comparison with the distribution of the full 564 E+As. Considering our selection criteria (brightest, high $S/N$ and then nearby), distributions of the full 564 and 35 galaxies in redshift, absolute magnitude and diameter are generally consistent. Such consistency convinces us that the 35 E+A sample are generally good representatives of the full 564 E+As.

\begin{table*}
 \centering
 \begin{minipage}{160mm}
  \caption{List of the 35 E+A sample selected from Goto's catalogue. Column 1 is the Galaxy Number (GN) we named for each of the sample. Column 2 is the SDSS ID. The rest of columns represent Petrosian magnitude in r-band, S/N ratio in r-band, redshift, right ascension (RA), declination (DEC) and EW of H${\rmn{\delta}}$, [O{\sc{ii}}] and H${\rmn{\alpha}}$, respectively. It should be noted that EW values here are measured by \citet{b42}.}
  \begin{tabular}{@{}cccccccccc@{}}
  \hline
   GN &IAU            & Petro  &$S/N$  &z  &RA  &DEC  &EW                 &EW             &EW\\
       &designation    & mag(r) &(r)    &   &    &     &H${\rmn{\delta}}$  &[O{\sc{ii}}]  &H${\rmn{\alpha}}$\\
   
 \hline
 1& SDSS $J$130525.82$+$533530.3 &14.49 &50.4  &0.0381 &196.35761 &53.591759 &5.06 &1.12  &2.38\\
 2& SDSS $J$162702.55$+$432833.9 &14.64 &45.0  &0.0463 &246.76067 &43.476095 &5.68 &-1.42 &0.68\\
 3& SDSS $J$123936.05$+$122619.9 &14.69 &43.4  &0.0409 &189.90021 &12.438888 &5.76 &0.86  &1.46 \\
 4& SDSS $J$120523.24$+$643029.6 &14.93 &41.4  &0.0822 &181.34687 &64.508254 &6.79 &0.90  &1.78 \\
 5& SDSS $J$115821.65$+$363820.4 &14.97 &48.3  &0.0656 &179.59021 &36.639018 &5.71 &1.41  &2.49 \\
 6& SDSS $J$145455.44$+$453126.5 &15.04 &40.6  &0.0366 &223.73102 &45.524055 &5.16 &0.96  &1.00 \\
 7& SDSS $J$155435.53$+$291319.9 &15.22 &45.7  &0.0943 &238.64808 &29.222199 &5.89 &-0.60 &1.29\\
 8& SDSS $J$080218.39$+$323207.8 &15.22 &45.1  &0.0382 &120.57664 &32.535504 &5.54 &-1.75 &1.54 \\
 9& SDSS $J$141419.29$-$031111.5 &15.24 &38.9  &0.0471 &213.58040 &-3.1865321&5.82 &-1.12 &0.32 \\
 10& SDSS $J$084918.88$+$462252.5 &15.30 &47.4  &0.0410 &132.32870 &46.381272 &8.34 &-1.16 &1.91\\
 11& SDSS $J$120419.07$-$001855.8 &15.30 &40.1  &0.0938 &181.07948 &-0.31550661&7.98&-1.18 &2.09\\
 12& SDSS $J$105220.44$+$054941.5 &15.34 &46.1  &0.0412 &163.08521 &5.8282184 &6.21 &-0.63 &1.92 \\
 13& SDSS $J$161330.18$+$510335.5 &15.38 &33.9  &0.0336 &243.37578 &51.059881 &7.57 &1.06 &0.55 \\
 14& SDSS $J$210258.87$+$103300.6 &15.42 &49.9  &0.0928 &315.74529 &10.550177 &5.19 &0.30 &1.81\\
 15& SDSS $J$100743.62$+$554934.5 &15.43 &54.5  &0.0452 &151.93174 &55.826261 &5.91 &-2.06 &1.58\\
 16& SDSS $J$233712.77$-$105800.4 &15.43 &44.4  &0.0783 &354.30318 &-10.966755 &6.66 &0.05 &1.76 \\
 17& SDSS $J$125820.71$+$613039.4 &15.49 &40.6  &0.0905 &194.58634 &61.510986 &7.28 &-1.15 &-1.81\\
 18& SDSS $J$153016.06$+$373346.1 &15.57 &53.1  &0.0775 &232.56693 &37.562823 &7.10 &-1.27 &-2.82\\
 19& SDSS $J$124204.66$+$150905.6 &15.59 &43.9  &0.0872 &190.51944 &15.151574 &7.40 &0.63 &2.10\\
 20& SDSS $J$110540.70$+$055954.2 &15.64 &49.8  &0.0543 &166.41961 &5.9984044 &6.89 &-1.42 &1.01\\
 21& SDSS $J$113040.84$+$562910.0 &15.65 &52.9  &0.0622 &172.67019 &56.486122 &5.60 &-0.98 &2.16\\
 22& SDSS $J$104220.14$+$564855.7 &15.68 &39.5  &0.0909 &160.58393 &56.815500 &5.50 &-1.78 &1.18\\
 23& SDSS $J$080925.07$+$305652.8 &15.70 &44.7  &0.0836 &122.35451 &30.948015 &5.48 &-1.86 &0.83\\
 24& SDSS $J$133757.98$+$654410.5 &15.76 &42.3  &0.0665 &204.49158 &65.736242 &8.02 &-1.36 &2.29\\
 25& SDSS $J$115837.72$-$021710.9 &15.79 &37.6  &0.0880 &179.65718 &-2.2863741 &6.08 &0.68 &1.75\\
 26& SDSS $J$133323.78$+$532130.1 &15.80 &33.4  &0.1023 &203.34913 &53.358373 &6.16 &-0.42 &1.06\\
 27& SDSS $J$083454.79$+$251258.5 &15.82 &38.8  &0.1322 &128.72832 &25.216253 &6.09 &1.04 &1.47\\
 28& SDSS $J$104204.13$+$293323.4 &15.85 &42.7  &0.0398 &160.51723 &29.556503 &6.44 &-1.03 &1.52\\
 29& SDSS $J$105743.93$+$123539.9 &15.91 &43.0  &0.1195 &164.43308 &12.594419 &8.36 &1.29 &2.48\\
 30& SDSS $J$104215.17$+$140636.3 &15.96 &53.9  &0.0543 &160.56321 &14.110087 &6.33 &-0.46 &2.16\\
 31& SDSS $J$162938.26$+$302924.9 &15.96 &36.8  &0.0971 &247.40944 &30.490256 &5.72 &-0.69 &1.51\\
 32& SDSS $J$092006.43$+$015807.7 &15.97 &50.5  &0.0850 &140.02682 &1.9687923 &6.01 &-0.54 &-1.52\\
 33& SDSS $J$134802.18$+$020405.6 &15.97 &38.6  &0.0678 &207.00909 &2.0682588 &6.66 &-0.02 &2.11\\
 34& SDSS $J$124534.16$+$402559.1 &15.98 &43.0  &0.0822 &191.39237 &40.433093 &6.18 &0.80 &1.70\\
 35& SDSS $J$135030.77$+$012804.7 &15.99 &39.1  &0.0728 &207.62819 &1.4679767 &7.12 &-1.70 &0.63\\
\hline
\end{tabular}
\end{minipage}
\end{table*}

\begin{figure*}
\centering
 \includegraphics[scale=0.9]{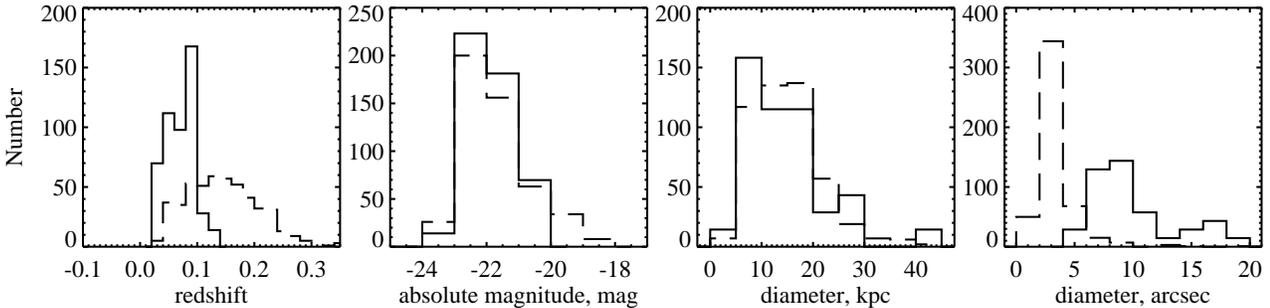}
 \caption[]{Comparison of distribution histograms of some major properties of E+As. 
We compared the redshift, absolute magnitude in mag, diameters in kpc,
and in arcsec for the 35 E+A sample (solid lines) and the full 564 E+A galaxies 
(dashed lines). We use bin=0.02, 1.0 mag, 5.0 kpc, and 2.0 arcsec, respectively. For the 35 E+A sample, the mean values of redshifts, absolute magnitudes in mag, and diameters in kpc and in arcsec are respectively 0.0705, -21.86 mag, 14.27 kpc and 11.27 arcsec. For the full 564 E+A galaxies, they are respectively 0.1414, -21.73 mag, 15.18 kpc and 3.34 arcsec.  }
\end{figure*}

\section[]{Constructing stellar populations for our sample}
The SFHs are always likely irregular, since they may be both triggered and truncated by environmental effects, which have some randomness\citep{b3,b4,b5}. For example, E+A galaxies started violent starbursts recently and stopped starbursts abruptly \citep{b32,b33,b34}. For exploring SFHs of galaxies, it is essential to study the stellar populations of galaxies which can be constructed by analysing the galaxies spectra. The average properties of all the stellar populations in a galaxy can be simply derived by 1-population (a single stellar population-a SSP) fit for the galaxy spectrum, and the obtained average properties are commonly described by SSP-equivalent age and metallicity.
  
In this section, we fit each of the sample with 1-population model to derive the equivalent properties of the representative stellar population for each of the sample. The principle of our analysis method is to compare an observed spectrum with a linear combination of some non-linear components (SSPs). We adopt two independent stellar population models in this work. One is the Pegase-HR/Elodie3.1 population model, and the other is the Vazdekis/Miles population model.

The Pegase-HR/Elodie3.1 (PE) population model uses the Elodie 3.1 stellar library \citep{b7,b8}, which is the newest version of the Elodie library, containing 1962 spectra of 1388 stars from the spectroscopic archive of Observatoire de Haute-Provence \citep{b16}. The wavelength coverage is 3900 $\sim$ 6800$\rmn{\AA}$, and the spectral resolution (FWHM) is 0.55 $\rmn{\AA}$. The principle of the data reduction, flux calibration and determination of the atmospheric parameters are described in \citet{b7}. Then the library is operated by the Pegase-HR code\citep{b6}, which allows a choice of different physical ingredients (initial mass function-IMF, evolutionary track, SFR). The isochrones are solar-scaled at different values of the total metallicity Z. The evolutionary tracks are taken from Padova 94 \citep{b43}, and the SSPs are computed with Salpeter IMF (\citet{b44}; 0.1 $M_{\rmn{\sun}}$ $\lid$ M $\lid$ 125 $M_{\rmn{\sun}}$). Ultimately, the PE model consists of 476 SSPs, covering 0.01 $\sim$ 20 Gyr for age and -2.30 $\sim$ 0.70 dex for [Fe/H]. 

The Vazdekis/Miles (VM) population model advances the previous model in \citet{b9} and \citet{b10}. It uses the Miles library \citep{b17}, which has a 2.3 $\rmn{\AA}$ spectral resolution (FWHM) and 3525 $\sim$ 7500 $\rmn{\AA}$ wavelength range. This library is believed to be better flux calibrated than any other. Padova 2000 isochrones \citep{b45} are used, which has hotter red giant branch, and the SSPs are computed with Salpeter IMF \citep{b44}. The generated VM model consists of 276 SSPs, covering 0.1 $\sim$ 17.7 Gyr for age and -1.7 $\sim$ 0.2 dex for [Fe/H]. 

We accomplish our analysis by ULySS \citep{b11}, which is an open-source software package enabling full-spectrum fit for the study of stellar populations of galaxies and star clusters. In ULySS, an observed spectrum is fitted with a model expressed as a linear combination of non-linear components. A component is a non-linear function of age, [Fe/H], [Mg/Fe], and wavelength. The ULySS method minimizes $\chi^2$ value when matching an observed spectrum with models. It is worth mentioning that ULySS adopts the appoach of injecting the line spread function (LSF) into the model to match the resolution between the model and the observation. More details about ULySS have been presented in \citet{b11}.

\subsection{Pre-treating: Resolution matching}
To compare SSP model with an observed spectrum, we have to firstly match resolutions of the model and the observation since they
are different here.
The spectral resolution is characterized by the instrumental broadening or the line spread function (LSF). LSF for spectra is equivalent to the point spread function (PSF) for images. In practice, LSF is not necessarily a Gaussian and varies with wavelength. The details on LSF have been described in \citet{b18,b11}. To match the resolutions by ULySS, we need to determine the relative LSF between the model and the observation, and then inject this relative LSF into the model. Our sample is selected from SDSS, so we just need to choose any spectrum from SDSS observations.
Here we use the spectrum of M67, which is a composite spectrum of stars from SDSS observations in M67 clusters and already contained in ULySS package, as representative of SDSS observations. ULySS calculate the relative LSF by comparing this composite spectrum of M67 with that extracted from the PE model. Then we inject this relative LSF to the PE model to generate the resolution-matched PE model by the LSF convolution function in ULySS package. Similarly, for the VM model, we do the same steps as above to generate the resolution-matched VM model. In the following text, we will use the resolution-matched PE or VM model, although they are still named as PE model and VM model.

\subsection{A single stellar population for our sample}
The first step towards reconstructing the SFH of an object is to calculate the SSP-equivalent properties by fitting the spectrum with a SSP. This derived SSP represents the `luminosity-weighted' epoch of star formation. Therefore, the SSP-equivalent properties are corresponding to the `luminosity-weighted' average over the distributions and provide a general view of the stellar populations of the galaxy. 

In this section, we make a SSP model fit for each galaxy by using ULySS. Both the PE and VM model are independently used as population models. Finally, we obtain the SSP-equivalent age and metallicity for each galaxy.

We show the results of a SSP fit with the PE model in Table 3 and with the VM model in Table 4. In both tables, the errors on the parameters are the formal 1-$\sigma$ errors, computed from the covariance matrix by the ULySS calling function: MPFIT which is a widely used fitting function. MPFIT gives out the optimal parameters from the best fit and the 1-$\sigma$ errors on them. More details on the errors can be obtained from the MPFIT algorithm (http://cow.physics.wisc.edu/~craigm/idl/idl.html). For a comparison, we show distribution histograms of SSP-equivalent age, metallicity and velocity dispersion of the sample in Figure 2 (solid lines). Figure 2 shows that no matter which model is used for fit, the derived SSP-equivalent ages are concentrating on about 1 Gyr, the metallicities distribute mostly between -0.50 and +0.20 dex, and the velocity dispersions cover from 100 to 220 km/s. Such results might indicate that the populations aged around 1Gyr would be the 'light-weighted' components in E+A galaxies.

Then we make a SSP fit for the total 564 E+A galaxies from Goto's catalogue with both the PE and VM model, and we also present the distribution histograms of the results in Figure 2 (dot-dashed lines). We calculate the mean SSP-equivalent age, metallicity and velocity dispersion and tabulate them in Table 2 for clear comparisons.
 
The results from both models are not exactly the same. The differences might arise from the difference and independence of the PE and VM model. The SSPs of the two models are generated by different population synthesis codes: Pegase.HR and Vazdekis, different stellar libraries: ELODIE3.1 and Miles, and different choices of the physical ingredients: evolutionary tracks, etc. However, the comparison in Figure 2 and Table 2 shows that the results from the PE and VM model are generally consistent.

In addition, both Figure 2 and Table 2 show the consistency between our 35 E+A sample and the full 564 E+A galaxies in the aspect of the SSP-equivalent age, metallicity and velocity dispersion. This consistency further proves that our 35 E+A sample is a good representative of the full 564 E+A galaxies.
\begin{table}
  \caption{Comparison of the mean SSP-equivalent age ($M_{age}$), [Fe/H] ($M_{[Fe/H]}$) and velocity dispersion ($M_{\sigma}$) for the 35 brightest E+As and the 564 E+As, illustrated respectively from the PE and VM model .
            }
  \begin{tabular*}{84mm}{@{}l|rr|rr|rr@{}}
\hline
 Sample &\multicolumn{2}{|c}{$M_{age}$(Gyr)} &\multicolumn{2}{|c}{$M_{[Fe/H]}$(dex)} &\multicolumn{2}{|c}{$M_{\sigma}$(km/s)}\\
         &PE             &VM  & PE              &VM  & PE                            &VM\\
 \hline
35 E+As &0.87 &1.14 &-0.04 &-0.38 &159 &141\\
564 E+As&1.01 &1.28 &-0.11 &-0.38 &144 &129\\
\hline
\end{tabular*}
\end{table}

\begin{table}
  \caption{Results of a SSP fit with the PE model. 
            Column 1 is the Galaxy Number (GN).
            Column 2 is the $\chi^2$ value from the best fit. 
            Column 3 and 4 are respectively the SSP-equivalent age in Gyr, [Fe/H] in dex and the errors on them.
            }
  \begin{tabular*}{84mm}{@{}p{8mm}p{8mm}rr}
\hline
  GN & $\chi^2$ & SSP-equivalent age  & SSP-equivalent [Fe/H]\\
       &          & (Gyr)               & (dex)\\
 \hline
1	  	&0.43	&1.10$\pm$0.03	&0.02$\pm$0.03	\\
2	  	&0.64	&1.12$\pm$0.04	&-0.04$\pm$0.05	\\
3	  	&0.58	&1.09$\pm$0.04	&-0.02$\pm$0.05	\\
4	 	&0.67	&0.82$\pm$0.03	&0.16$\pm$0.04	\\
5	  	&0.44	&0.80$\pm$0.02	&0.12$\pm$0.03	\\
6	  	&0.61	&0.86$\pm$0.02	&0.17$\pm$0.03	\\
7	  	&0.59	&1.03$\pm$0.03	&-0.05$\pm$0.05	\\
8	        &0.51	&0.93$\pm$0.02	&0.14$\pm$0.03	\\
9	  	&0.65	&1.15$\pm$0.06	&-0.10$\pm$0.05	\\
10		&0.53	&0.67$\pm$0.02	&-0.24$\pm$0.04	\\
11		&0.69	&0.71$\pm$0.02	&-0.19$\pm$0.06	\\
12		&0.54	&0.99$\pm$0.03	&-0.07$\pm$0.04	\\
13		&0.66	&0.06$\pm$0.00	&-0.07$\pm$0.06	\\
14		&0.61	&1.17$\pm$0.05	&0.02$\pm$0.03	\\
15		&0.67	&1.00$\pm$0.03	&-0.12$\pm$0.04	\\
16		&0.53	&0.80$\pm$0.02	&0.09$\pm$0.04	\\
17		&0.69	&0.82$\pm$0.03	&-0.68$\pm$0.08	\\
18		&0.56	&0.74$\pm$0.02	&-0.06$\pm$0.04	\\
19		&0.55	&0.73$\pm$0.02	&0.03$\pm$0.04	\\
20		&0.53	&0.78$\pm$0.02	&-0.08$\pm$0.05	\\
21		&0.64	&0.87$\pm$0.02	&0.08$\pm$0.03	\\
22		&0.64	&1.13$\pm$0.05	&-0.07$\pm$0.05	\\
23		&0.58	&1.00$\pm$0.03	&-0.01$\pm$0.05	\\
24		&0.62	&0.71$\pm$0.02	&-0.11$\pm$0.05	\\
25		&0.65	&0.88$\pm$0.03	&0.06$\pm$0.04	\\
26		&0.72	&0.88$\pm$0.04	&-0.01$\pm$0.06	\\
27		&0.64	&0.91$\pm$0.03	&-0.02$\pm$0.06	\\
28		&0.59	&0.94$\pm$0.03	&-0.05$\pm$0.04	\\
29		&0.57	&0.67$\pm$0.02	&-0.20$\pm$0.05	\\
30		&0.65	&0.87$\pm$0.02	&-0.14$\pm$0.03	\\
31		&0.74	&1.06$\pm$0.05	&-0.06$\pm$0.06	\\
32		&0.91	&0.96$\pm$0.04	&-0.14$\pm$0.05	\\
33		&0.64	&0.83$\pm$0.03	&-0.09$\pm$0.05	\\
34		&0.62	&0.82$\pm$0.02	&0.11$\pm$0.04	\\
35		&0.63	&0.68$\pm$0.02	&0.12$\pm$0.04	\\
\hline
\end{tabular*}
\end{table}

\begin{table}
  \caption{Results of a SSP fit with the VM model.
            Column 1 is the Galaxy Number (GN).
            Column 2 is the $\chi^2$ value from the best fit. 
            Column 3 and 4 are respectively the SSP-equivalent age in Gyr, [Fe/H] in dex and the errors on them.}
  \begin{tabular*}{84mm}{@{}p{8mm}p{8mm}rr}
\hline
  GN & $\chi^2$ & SSP-equivalent age   & SSP-equivalent [Fe/H]  \\
       &         &  (Gyr)                &   (dex)\\
  
 \hline
1	  	&0.50   &1.39$\pm$0.05 &-0.23$\pm$0.01	\\
2	  	&0.63   &1.38$\pm$0.05 &-0.37$\pm$0.02	\\
3	  	&0.64   &1.39$\pm$0.06 &-0.38$\pm$0.02	\\
4	  	&0.76   &1.14$\pm$0.02 &-0.33$\pm$0.03	\\
5	  	&0.53 &1.02$\pm$0.02 &-0.26$\pm$0.03	\\
6	  	&0.67 &1.38$\pm$0.05 &-0.35$\pm$0.02	\\
7	  	&0.68 &1.38$\pm$0.05 &-0.35$\pm$0.01	\\
8	  	&0.59 &1.14$\pm$0.01 &-0.15$\pm$0.02	\\
9	  	&0.69 &1.38$\pm$0.07 &-0.29$\pm$0.02	\\
10		&0.57 &0.75$\pm$0.01 &-0.34$\pm$0.03	\\
11		&0.70 &0.84$\pm$0.02 &-0.38$\pm$0.03	\\
12		&0.57 &1.14$\pm$0.01 &-0.28$\pm$0.02	\\
13		&0.77 &0.60$\pm$0.01 &-1.33$\pm$0.06	\\
14		&0.70 &1.39$\pm$0.05 &-0.26$\pm$0.01	\\
15		&0.72 &0.96$\pm$0.01 &-0.05$\pm$0.02	\\
16		&0.60 &0.90$\pm$0.02 &-0.10$\pm$0.02	\\
17		&0.72 &1.49$\pm$0.02 &-1.21$\pm$0.03	\\
18		&0.73 &1.19$\pm$0.01 &-0.83$\pm$0.03	\\
19		&0.64 &0.89$\pm$0.01 &-0.28$\pm$0.02	\\
20		&0.62 &0.84$\pm$0.02 &-0.11$\pm$0.03	\\
21		&0.73 &0.86$\pm$0.02 &0.12$\pm$0.02	\\
22		&0.69 &1.38$\pm$0.06 &-0.33$\pm$0.02	\\
23		&0.62 &1.38$\pm$0.04 &-0.38$\pm$0.01	\\
24		&0.63 &0.80$\pm$0.01 &-0.21$\pm$0.02	\\
25		&0.76 &1.17$\pm$0.02 &-0.38$\pm$0.03	\\
26		&0.87 &1.15$\pm$0.02 &-0.50$\pm$0.03	\\
27		&0.74 &1.60$\pm$0.18 &-0.67$\pm$0.02	\\
28		&0.63 &1.16$\pm$0.01 &-0.35$\pm$0.03	\\
29		&0.66 &0.75$\pm$0.01 &-0.27$\pm$0.02	\\
30		&0.69 &0.96$\pm$0.02 &-0.31$\pm$0.02	\\
31		&0.70 &1.37$\pm$0.06 &-0.39$\pm$0.02	\\
32		&0.96 &1.87$\pm$0.06 &-0.73$\pm$0.03	\\
33		&0.73 &0.94$\pm$0.02 &-0.23$\pm$0.03	\\
34		&0.66 &1.14$\pm$0.01 &-0.38$\pm$0.03	\\
35		&0.67 &0.88$\pm$0.02 &-0.30$\pm$0.03	\\
\hline
\end{tabular*}
\end{table}

\begin{figure*}
\centering
 \includegraphics[scale=0.8]{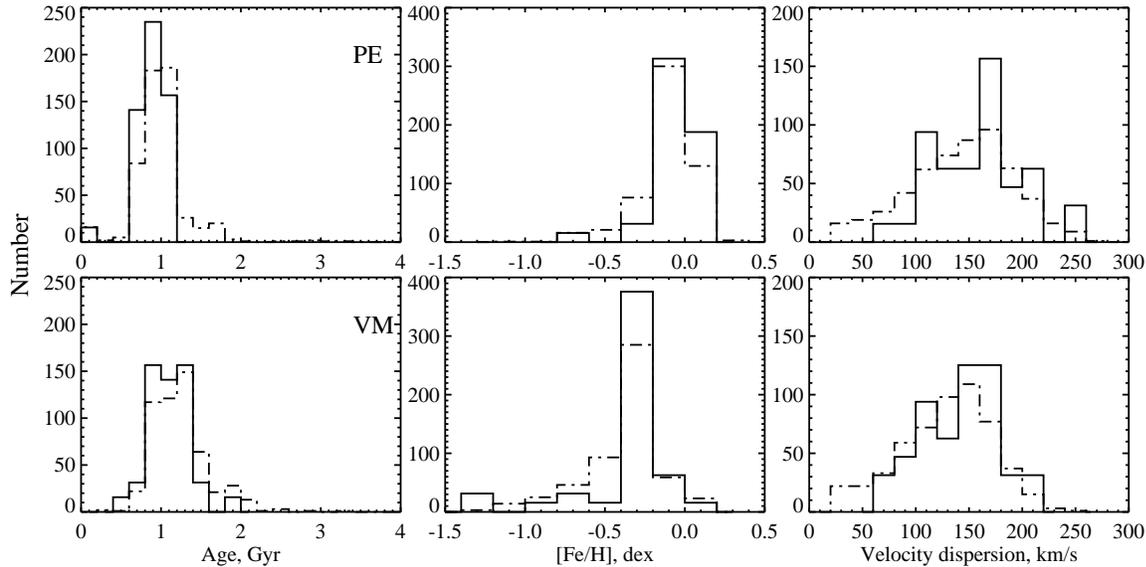}
 \caption[]{Parameters distributions from a SSP fit with the PE model and the VM model. The solid lines show the distribution histograms of the 35 E+A sample, and the dot-dashed lines are for the full 564 E+A galaxies. We present the SSP-equivalent age, metallicity and velocity dispersion distribution of them.  
The top panel shows the results from the PE model, and the bottom is from the VM model.
In the top panel, for the 35 E+A sample, the mean SSP-equivalent age, [Fe/H] and velocity dispersion are 0.87 Gyr, -0.04 dex and 159 km/s; for the 564 E+A galaxies, the mean values are respectively 1.01 Gyr, -0.11 dex and 144 km/s. In the bottom panel, for the 35 E+A sample, the mean values are respectively 1.14 Gyr, -0.38 dex and 141 km/s; for the 564 E+A galaxies, the mean values are respectively 1.28 Gyr, -0.38 dex and 129 km/s. These mean values are tabulated in Table 2 for a direct comparison.  }
\end{figure*}

\subsection{The Composite Stellar Population for our sample}
It is too simple to fit the spectra of galaxies only using a SSP, since the galaxies have generally a complex SFH. Retracing the SFR along the history is a fundamental piece of information to understand the physics of the galaxies. In principle, one can obtain such information by directly fitting the galaxy with a positive linear combination of many SSPs, which means a composite stellar population (CSP) fit for the galaxy, but such an approach would be unstable because of the degeneracies between the SSP components. To circumvent these degeneracies, ULySS takes effective measures. It starts with the simple physical assumptions, such as the presence of an old stellar population, then divide the time axis into intervals by setting limits in two or more intervals. More details on dealing with degeneracies by ULySS have been analysed in \citet{b11}. 

 In our work, we analyse the spectrum of each galaxy in terms of three epochs which means to fit each spectrum with 3 stellar populations: an old stellar population (OSP), an intermediate-age stellar population (ISP), and a young stellar population (YSP). They are respectively chosen from the following 3 boxes divided by different limits of age and metallicity. 

When we are using the PE model, we choose the boxes as follows:

(i) For the OSP box, we fix its age to 12Gyr and set the [Fe/H] free (we fix the age to maintain the degree of freedom at the minimum, and meanwhile to see whether such an old stellar component can be discovered in E+As).

(ii) For the ISP box, we make the age span a range from 1 to 5 Gyr and the [Fe/H] span from -1.00 to 0.20 dex (for several galaxies, the ISP component is pushing on such [Fe/H] limits, so we set the [Fe/H] free for these galaxies). 

(iii) For the YSP box, we make the age span a range from 0.01 to 1 Gyr and the [Fe/H] span from -1.00 to 0.70 dex.

We fit each of the sample with 3 populations (OSP + ISP + YSP) by using ULySS, which gives the optimal results from the best fit by adjusting the set of parameters. It gives the optimal age, [Fe/H], light fraction (LF), mass fraction (MF) and errors on them for each component. During the fit, the boxes are suitable for all the sample galaxies except for several ones which have their optimal paramters reaching to the boundaries of boxes. For such cases, we only give their upper or lower limits.

Results from this 3-population (3 SSPs) fit are generally better than those from 1-population fit ($\S$ 3.2) and 2-population (2 SSPs) fit (In $\S$ 6, we will additionally make the 2-SSP fit for each of the sample to compare E+As with ellipticals). The $\chi^2$s are lower and the residuals, especially at the absorption features, are smaller. We show the exemplary fit and the residuals of Galaxy number 1 as a representative in Figure 3. Table 5 presents the optimal parameters and errors on them from this 3-SSP fit with the PE model, and the optimal age, [Fe/H], LF and MF are tabulated for each component. The errors on the optimal age and [Fe/H] in Table 5 are obtained from the MPFIT fitting process similarly as mentioned for errors in Table 3, and errors on the light fraction and mass fraction are determined by a multiple linear regression routine called by ULySS. 

In Table 5, we find that 12Gyr-old components can be detected in all of the 35 galaxies. Very young($< $0.13 Gyr) components are discoverd in 3 of the sample (Galaxy number 3, 13 and 17), which indicates a quite recent starburst in the three E+A galaxies. For the majority of the sample (23 out of 35), YSPs dominate the light, since they have much larger light fraction than ISPs and OSPs. For 10 of the remaining 12 galaxies, the ISPs dominate the light. For the mass, all the 35 galaxies except 2 are dominated by the OSPs, since they have much higher mass fraction. To sumarize, most of E+As are dominated by YSPs in light and some are dominated by ISPs in light. While the OSPs are not the dominant contributors to light, but they are the dominant contributors to mass, statistically. This conclusion can be obtained more clearly from Figure 4 and Figure 5, which will be described later.
 
For comparison we repeat fitting each of the sample with 3 SSPs from the VM model. To adjust to the narrower coverage of the VM model, we change the limits for boxes as follows:

(i) For the OSP box, we still fix its age to 12Gyr and set the [Fe/H] free (we fix the age to maintain the degree of freedom at the minimum, and meanwhile to see whether such an old stellar component can be discovered in E+As).

(ii) For the ISP box, we make the age span a range from 1 to 5 Gyr and the [Fe/H] span from -1.00 to 0.10 dex (Here we choose 0.10 dex because of the small upper limit (0.20 dex) of the VM model for [Fe/H]; for several galaxies, we set the [Fe/H] free since the ISP component is pushing on the [Fe/H] limits).  

(iii) For the YSP box, we make the age span a range from 0.1 to 1 Gyr and the
[Fe/H] span from -1.00 to 0.20 dex (no younger populations than 0.1Gyr and no richer populations than 0.20 dex are contained in the VM model).

In Table 6, we list the results for each component from the 3-population fit with the VM model. All of the sample galaxies have 12Gyr-old populations. Similarly as in Table 5, very young populations ($<$ 0.13) are detected in the same three galaxies (Galaxy number 3, 13 and 17). All of the results and conclusions from 3-population fit with the VM model (Table 6) have a good agreement with those from the PE model (Table 5). This general agreement could also be seen from Figure 4 and Figure 5, which will be described later.

For a clear comparison of the results of the 3-population fit with the two independent population models, we show the distribution histograms of the light fraction (Figure 4) and mass fraction (Figure 5) of the YSP, ISP and OSP for all of the 35 E+A sample. In both figures, the top panel shows results from the PE model, and the bottom panel shows results from the VM model. Light fraction distributions of the YSP, ISP and OSP are presented from left to right, respectively. Obviously, we can conclude that no matter which model is used for fit, in a statistical sense, YSPs contribute the most to the light, then the ISPs, and finally the OSPs; however, OSPs contribute the most to the mass, then the ISPs, and finally the YSPs. This conclusion can be obtained more clearly from Table 7, which presents the mean light contributions, mass contributions and ages of YSP, ISP and OSP by using the PE and VM models. YSPs contribute about 50$\%$ of the light, and ISPs and OSPs contribute another 50$\%$. However, OSPs contribute about 70$\%$ of the mass, and ISPs and OSPs are presented in the rest 30$\%$.
For the mean age, YSP of the sample is concentrating on a mean age of 0.54 Gyr with the PE model and 0.46 Gyr with the VM model, the ISPs are concentrating on the mean age of 1.82 Gyr with the PE model and 1.82 Gyr with the VM model. The two independent models give generally consistent results.


\begin{table*}
 \caption{Parameters from 3-population (OSP + ISP + YSP) fit with the PE model for each of the sample. This table presents the population histories of the 35 E+As, reconstructed from 3 episodes of star formation. Column 1 is the Galaxy Number (GN) for the E+A sample listed in Table 1. Column 2 is $\chi^2$ value from the best fit. The following are age (in Gyr), [Fe/H] (in dex), mass fraction (MF, in $\%$), and light fraction (LF, in $\%$) for each of the component (OSP, ISP, and YSP). The last line lists the results of the combined spectrum, which will be described later in $\S$ 4.2.
          }
\setlength{\tabcolsep}{1.0pt}
\renewcommand{\arraystretch}{1.1}
\begin{tabular*}{176mm}{@{\extracolsep{\fill}}lc|ccc|cccc|cccc}
\hline
GN&$\chi^2$  & \multicolumn{3}{|c}{OSP (fixed to 12Gyr)} & \multicolumn{4}{|c}{ISP}         & \multicolumn{4}{|c}{YSP}  \\
\cline{3-13}
      &         & [Fe/H]&MF     &LF                    & Age & [Fe/H]&MF    & LF     & Age & [Fe/H] &MF    & LF\\
      &         &(dex)  &($\%$) &($\%$)                &(Gyr)&(dex)  &($\%$)&($\%$)  &(Gyr)&(dex)   &($\%$)&($\%$)\\
  
\hline\hline

1& 0.40& -0.22$\pm$0.64& 61.16$\pm$3.28& 16.46$\pm$0.88& 1.05$\pm$0.29& -0.50$\pm$0.37& 14.69$\pm$0.24& 43.65$\pm$0.72& 0.99$\pm$0.33&
0.52$\pm$0.66& 24.15$\pm$0.95& 38.89$\pm$1.57\\
2& 0.58& 0.21$\pm$0.11& 92.51$\pm$1.54& 38.80$\pm$0.64& 1.47$\pm$0.32& -0.69$\pm$0.25& 4.72$\pm$0.32& 28.53$\pm$1.91& 0.51$\pm$0.19&
-0.38$\pm$0.28& 2.77$\pm$0.11& 32.67$\pm$1.28\\
3& 0.51& 0.25$\pm$0.08& 94.79$\pm$0.57& 42.27$\pm$0.25& 0.73$\pm$0.18& -0.38$\pm$0.12& 4.95$\pm$0.08& 48.26$\pm$0.82& 0.09$\pm$0.09& 0.30$\pm$0.77&
0.26$\pm$0.02& 9.47$\pm$0.59\\
4& 0.63& 0.26$\pm$0.18& 89.40$\pm$0.52& 28.11$\pm$0.16& 1.21$\pm$0.76& -1.00$\pm$0.63& 2.54$\pm$0.39& 15.25$\pm$2.35& 0.58$\pm$0.10& 0.06$\pm$0.23&
8.06$\pm$0.32& 56.64$\pm$2.24\\
5& 0.41& 0.21$\pm$0.61& 56.91$\pm$5.51& 9.38$\pm$0.91& 1.49$\pm$0.47& 0.30$\pm$0.20& 28.04$\pm$1.32& 32.81$\pm$1.54& 0.60$\pm$0.07& -0.23$\pm$0.10&
15.05$\pm$0.17& 57.81$\pm$0.65\\
6& 0.57& 0.22$\pm$0.25& 61.76$\pm$6.75& 12.08$\pm$1.32& 1.48$\pm$0.27& 0.41$\pm$0.14& 24.80$\pm$1.65& 32.67$\pm$2.17& 0.70$\pm$0.07&
-0.25$\pm$0.13& 13.45$\pm$0.21& 55.25$\pm$0.87\\
7& 0.57& 0.29$\pm$0.29& 80.51$\pm$0.97& 19.27$\pm$0.23& 1.55$\pm$0.27& -0.84$\pm$0.36& 11.69$\pm$0.54& 42.19$\pm$1.95& 0.56$\pm$0.07&
$\gid$ 0.30& 7.80$\pm$0.36& 38.54$\pm$1.76\\
8& 0.50& -0.06$\pm$0.19& 63.45$\pm$1.01& 13.19$\pm$0.21& 1.42$\pm$0.37& -0.70$\pm$0.59& 8.89$\pm$0.78& 21.14$\pm$1.86& 0.73$\pm$0.04&
$\gid$ 0.30& 27.66$\pm$0.83& 65.67$\pm$1.96\\
9& 0.64& 0.36$\pm$0.22& 80.26$\pm$3.28& 19.23$\pm$0.79& 1.27$\pm$0.12& -0.71$\pm$0.47& 14.47$\pm$0.49& 64.66$\pm$2.18& $\gid$1.00&
0.30$\pm$0.52& 5.27$\pm$0.95& 16.12$\pm$2.92\\
10& 0.47& -0.19$\pm$0.45& 76.20$\pm$7.79& 21.30$\pm$2.18& 6.66$\pm$8.75& 0.42$\pm$0.93& 13.34$\pm$6.60& 3.87$\pm$1.91& 0.33$\pm$0.03&
-0.14$\pm$0.08& 10.46$\pm$0.04& 74.84$\pm$0.29\\
11& 0.61& 0.13$\pm$0.43& 74.10$\pm$6.31& 18.66$\pm$1.59& 3.89$\pm$1.62& -0.12$\pm$0.71& 17.80$\pm$2.68& 14.25$\pm$2.15& 0.34$\pm$0.05&
-0.10$\pm$0.15& 8.11$\pm$0.07& 67.08$\pm$0.58\\
12& 0.52& 0.25$\pm$0.45& 49.78$\pm$12.80& 7.45$\pm$1.92& 1.84$\pm$0.94& 0.52$\pm$0.14& 18.80$\pm$3.97& 13.53$\pm$2.86& 0.96$\pm$0.06&
-0.31$\pm$0.10& 31.42$\pm$0.38& 79.02$\pm$0.96\\
13& 0.66& -0.69$\pm$1.14& 51.97$\pm$9.16& 8.88$\pm$1.57& 7.87$\pm$5.97& 0.69$\pm$1.17& 37.90$\pm$12.95& 4.67$\pm$1.60& 0.08$\pm$0.01&
-0.20$\pm$0.09& 10.12$\pm$0.02& 86.45$\pm$0.17\\
14& 0.55& 0.15$\pm$0.16& 79.83$\pm$3.15& 26.56$\pm$1.05& 1.38$\pm$0.29& 0.49$\pm$0.12& 13.90$\pm$0.71& 30.04$\pm$1.54& 0.75$\pm$0.10&
-0.59$\pm$0.20& 6.27$\pm$0.07& 43.41$\pm$0.51\\
15& 0.64& 0.54$\pm$0.11& 69.86$\pm$2.05& 10.43$\pm$0.31& 1.28$\pm$0.24& -0.55$\pm$0.46& 17.65$\pm$1.05& 50.10$\pm$2.98& 0.74$\pm$0.34&
0.04$\pm$0.30& 12.49$\pm$0.86& 39.47$\pm$2.71\\
16& 0.50& 0.22$\pm$0.17& 87.77$\pm$1.30& 25.71$\pm$0.38& $\lid$1.00& -0.48$\pm$1.53& 1.93$\pm$1.11& 9.61$\pm$5.52& 0.64$\pm$0.33&
-0.14$\pm$0.40& 10.30$\pm$0.82& 64.68$\pm$5.16\\
17& 0.57&$\lid$-2.30& 18.09$\pm$1.40& 12.88$\pm$0.30& $\gid$5.00& 0.44$\pm$0.12& 78.98$\pm$0.32& 19.45$\pm$1.14& 0.12$\pm$0.01&
-0.95$\pm$0.22& 4.93$\pm$0.03&69.67$\pm$0.86\\
18& 0.51& 0.63$\pm$0.10& 81.42$\pm$1.78& 14.34$\pm$0.31& $\lid$1.00& -0.31$\pm$0.09& 16.05$\pm$0.20& 54.89$\pm$0.69& 0.15$\pm$0.03&
0.30$\pm$0.19& 2.53$\pm$0.03& 30.77$\pm$0.40\\
19& 0.53& 0.48$\pm$0.09& 80.31$\pm$2.73& 14.33$\pm$0.49& $\lid$1.00& 0.01$\pm$0.19& 7.85$\pm$0.49& 23.25$\pm$1.44& 0.59$\pm$0.08&
-0.27$\pm$0.09& 11.84$\pm$0.18& 62.42$\pm$0.97\\
20& 0.51& 0.32$\pm$0.69& 45.91$\pm$12.77& 4.88$\pm$1.36& 1.79$\pm$0.74& $\gid$0.70& 14.25$\pm$3.58& 6.72$\pm$1.69& 0.78$\pm$0.06&
-0.24$\pm$0.09& 39.85$\pm$0.16& 88.41$\pm$0.36\\
21& 0.61& 0.30$\pm$0.22& 72.08$\pm$0.72& 12.60$\pm$0.13& $\lid$1.00& -0.24$\pm$0.22& 19.96$\pm$0.87& 58.47$\pm$2.56& 0.53$\pm$0.25&
0.39$\pm$0.21& 7.96$\pm$0.69& 28.93$\pm$2.50\\
22& 0.59& -0.21$\pm$0.33& 78.24$\pm$4.36& 29.17$\pm$1.63& 1.28$\pm$0.54& 0.53$\pm$0.16& 14.88$\pm$1.13& 27.18$\pm$2.06& 0.62$\pm$0.18&
-0.49$\pm$0.24& 6.87$\pm$0.07& 43.65$\pm$0.47\\
23& 0.55& -0.40$\pm$0.21& 75.78$\pm$3.47& 26.55$\pm$1.22& 1.16$\pm$0.33& 0.37$\pm$0.18& 13.19$\pm$0.90& 25.02$\pm$1.71& 0.60$\pm$0.10&
-0.03$\pm$0.17& 11.04$\pm$0.12& 48.43$\pm$0.54\\
24& 0.57& 0.37$\pm$0.27& 76.08$\pm$8.91& 11.80$\pm$1.38& 1.29$\pm$0.58& 0.70$\pm$0.37& 4.18$\pm$2.11& 4.20$\pm$2.12& 0.61$\pm$0.05& -0.28$\pm$0.07&
19.74$\pm$0.18& 84.00$\pm$0.76\\
25& 0.61& 0.29$\pm$0.44& 54.79$\pm$7.03& 7.64$\pm$0.98& 1.21$\pm$0.32& 0.52$\pm$0.09& 19.25$\pm$1.63& 22.39$\pm$1.89& 0.79$\pm$0.07&
-0.16$\pm$0.10& 25.96$\pm$0.35& 69.97$\pm$0.93\\
26& 0.68& 0.36$\pm$0.14& 90.56$\pm$1.70& 27.22$\pm$0.51& 1.23$\pm$0.21& -0.56$\pm$0.30& 7.11$\pm$0.28& 38.36$\pm$1.50& 0.26$\pm$0.10&
-0.07$\pm$0.39& 2.33$\pm$0.07& 34.42$\pm$1.01\\
27& 0.58& 0.18$\pm$0.12& 93.24$\pm$3.59& 38.64$\pm$1.49& 1.37$\pm$0.47& $\gid$0.70& 2.04$\pm$0.96& 4.24$\pm$1.99& 0.49$\pm$0.09&
-0.53$\pm$0.15& 4.72$\pm$0.04& 57.12$\pm$0.52\\
28& 0.55& 0.28$\pm$0.26& 74.91$\pm$3.04& 13.78$\pm$0.56& 1.38$\pm$0.16& -0.53$\pm$0.15& 17.09$\pm$0.76& 48.99$\pm$2.18& 0.48$\pm$0.16&
0.18$\pm$0.27& 8.00$\pm$0.35& 37.24$\pm$1.64\\
29& 0.54& 0.31$\pm$0.28& 84.87$\pm$8.75& 16.04$\pm$1.65& 1.86$\pm$3.84& 0.67$\pm$1.74& 0.26$\pm$2.55& 0.20$\pm$2.03& 0.51$\pm$0.05& -0.25$\pm$0.10&
14.87$\pm$0.07& 83.75$\pm$0.40\\
30& 0.63& -0.12$\pm$0.31& 60.50$\pm$1.88& 10.53$\pm$0.33& $\lid$1.00& -0.25$\pm$0.06& 36.83$\pm$0.39& 71.34$\pm$0.76& 0.16$\pm$0.03&
$\gid$0.30& 2.66$\pm$0.07& 18.13$\pm$0.45\\
31& 0.60& 0.21$\pm$0.14& 91.49$\pm$1.07& 35.99$\pm$0.42& 1.00$\pm$0.60& -0.63$\pm$0.35& 7.46$\pm$0.44& 52.18$\pm$3.05& 0.44$\pm$0.72&
-0.04$\pm$1.90& 1.06$\pm$0.24& 11.82$\pm$2.65\\
32& 0.86& 0.31$\pm$0.15& 81.71$\pm$0.59& 21.11$\pm$0.15& 3.91$\pm$1.84& $\lid$-2.30& 6.29$\pm$1.21& 14.02$\pm$2.69& 0.84$\pm$0.10&
-0.40$\pm$0.16& 11.99$\pm$0.48& 64.87$\pm$2.59\\
33& 0.63& 0.24$\pm$0.20& 79.36$\pm$2.21& 16.37$\pm$0.46& 1.05$\pm$0.27& -0.44$\pm$0.28& 12.75$\pm$0.97& 43.37$\pm$3.32& 0.55$\pm$0.26&
-0.09$\pm$0.52& 7.89$\pm$0.57& 40.26$\pm$2.88\\
34& 0.58& 0.27$\pm$0.15& 88.76$\pm$1.07& 26.96$\pm$0.33& 0.90$\pm$0.12& -0.29$\pm$0.13& 9.97$\pm$0.22& 54.75$\pm$1.19& 0.21$\pm$0.09&
0.33$\pm$0.42& 1.28$\pm$0.06& 18.29$\pm$0.89\\
35& 0.60& 0.19$\pm$0.50& 72.40$\pm$8.02& 15.09$\pm$1.67& 2.05$\pm$0.97& 0.23$\pm$0.73& 13.61$\pm$2.40& 14.59$\pm$2.57& 0.53$\pm$0.08&
-0.09$\pm$0.16& 13.99$\pm$0.18& 70.32$\pm$0.92\\
\hline
c& 0.22& -0.20$\pm$0.41& 57.95$\pm$6.64& 13.05$\pm$1.50& 1.35$\pm$0.73& 0.39$\pm$0.23& 22.44$\pm$1.74& 26.80$\pm$2.08& 0.72$\pm$0.09&
-0.22$\pm$0.15& 19.61$\pm$0.20& 60.14$\pm$0.62\\

\hline
\end{tabular*}
\end{table*}


\begin{table*}
 \caption{Parameters from 3-population (OSP + ISP + YSP) fit with the VM model for each of the sample. This table presents the population histories of the 35 E+As, reconstructed from 3 episodes of star formation. Column 1 is the Galaxy Number (GN) for the E+A sample listed in Table 1. Column 2 is $\chi^2$ value from the best fit. The following are age (in Gyr), [Fe/H] (in dex), mass fraction (MF, in $\%$), and light fraction (LF, in $\%$) for each of the component (OSP, ISP, and YSP).
          }
\setlength{\tabcolsep}{1.0pt}
\renewcommand{\arraystretch}{1.1}
\begin{tabular*}{174mm}{@{\extracolsep{\fill}}lc|ccc|cccc|cccc}
\hline
  GN&$\chi^2$  & \multicolumn{3}{|c}{OSP (fixed to 12Gyr)} & \multicolumn{4}{|c}{ISP}         & \multicolumn{4}{|c}{YSP}  \\
\cline{3-13}
      &         & [Fe/H]&MF     &LF                    & Age & [Fe/H]&MF    & LF     & Age & [Fe/H] &MF    & LF\\
      &         &(dex)  &($\%$) &($\%$)                &(Gyr)&(dex)  &($\%$)&($\%$)  &(Gyr)&(dex)   &($\%$)&($\%$)\\
  
\hline\hline

1& 0.40& $\gid$0.20& 73.62$\pm$2.91& 18.68$\pm$0.74& 1.27$\pm$0.21& -0.28$\pm$0.23& 7.07$\pm$1.10& 21.57$\pm$3.35& 0.87$\pm$0.09&
0.09$\pm$0.05& 19.31$\pm$0.85& 59.75$\pm$2.63\\
2& 0.56& 0.13$\pm$0.08& 93.23$\pm$1.42& 44.21$\pm$0.67& 1.42$\pm$0.63& -1.19$\pm$0.32& 2.55$\pm$0.28& 19.15$\pm$2.11& 0.60$\pm$0.09&
-0.06$\pm$0.14& 4.22$\pm$0.17& 36.64$\pm$1.46\\
3& 0.52& $\gid$0.20& 95.19$\pm$0.71& 47.10$\pm$0.35& 0.62$\pm$0.06& -0.17$\pm$0.10& 4.48$\pm$0.13& 43.09$\pm$1.25& 0.12$\pm$0.08&
0.20$\pm$0.58& 0.33$\pm$0.03& 9.81$\pm$0.92\\
4& 0.65& 0.05$\pm$0.14& 82.89$\pm$2.61& 27.04$\pm$0.85& $\lid$1.00& -0.05$\pm$0.15& 12.97$\pm$0.67& 44.02$\pm$2.28& 0.42$\pm$0.07&
0.19$\pm$0.12& 4.14$\pm$0.21& 28.95$\pm$1.44\\
5& 0.45& 0.20$\pm$-0.00& 8.27$\pm$8.68& 0.89$\pm$0.93& 1.61$\pm$0.13& 0.19$\pm$0.04& 72.84$\pm$1.93& 53.76$\pm$1.43& 0.49$\pm$0.03& -0.02$\pm$0.05&
18.89$\pm$0.21& 45.36$\pm$0.51\\
6& 0.57& 0.18$\pm$0.11& 80.58$\pm$4.98& 22.88$\pm$1.41& 1.34$\pm$0.15& -0.20$\pm$0.00& 10.35$\pm$1.09& 30.55$\pm$3.21& 0.55$\pm$0.03&
0.16$\pm$0.05& 9.07$\pm$0.35& 46.56$\pm$1.81\\
7& 0.62& -1.13$\pm$0.18& 4.56$\pm$4.74& 1.19$\pm$1.24& 1.60$\pm$0.16& $\gid$0.20& 69.56$\pm$1.49& 51.86$\pm$1.11& 0.67$\pm$0.04&
-0.03$\pm$0.07& 25.88$\pm$0.13& 46.95$\pm$0.23\\
8& 0.50& 0.10$\pm$0.18& 59.46$\pm$2.58& 11.71$\pm$0.51& 1.60$\pm$0.92& -0.53$\pm$0.49& 6.08$\pm$1.15& 11.14$\pm$2.11& 0.84$\pm$0.05& 0.14$\pm$0.04&
34.46$\pm$0.73& 77.15$\pm$1.64\\
9& 0.67& -0.95$\pm$0.24& 51.38$\pm$5.17& 19.87$\pm$2.00& 1.58$\pm$0.21& $\gid$0.20& 37.57$\pm$1.74& 45.04$\pm$2.08& 0.63$\pm$0.11&
-0.14$\pm$0.18& 11.04$\pm$0.09& 35.09$\pm$0.30\\
10& 0.49& -1.09$\pm$0.20& 35.05$\pm$2.06& 18.36$\pm$1.08& $\gid$5.00& $\gid$0.20& 55.80$\pm$2.51& 16.93$\pm$0.76& 0.29$\pm$0.01&
0.14$\pm$0.04& 9.15$\pm$0.05& 64.71$\pm$0.33\\
11& 0.61& $\gid$0.20& 70.68$\pm$6.08& 15.85$\pm$1.36& 2.75$\pm$0.76& -0.37$\pm$0.14& 22.21$\pm$1.66& 26.56$\pm$1.99& 0.25$\pm$0.02&
0.20$\pm$0.10& 7.11$\pm$0.08& 57.59$\pm$0.64\\
12& 0.52& -0.17$\pm$0.32& 33.11$\pm$10.73& 5.73$\pm$1.85& 1.49$\pm$0.09& 0.14$\pm$0.07& 45.80$\pm$2.90& 45.84$\pm$2.90& 0.68$\pm$0.06&
-0.14$\pm$0.08& 21.08$\pm$0.47& 48.43$\pm$1.07\\
13& 0.70& $\gid$0.20& 41.64$\pm$9.85& 3.50$\pm$0.83& $\gid$5.00& -1.54$\pm$0.35& 43.80$\pm$4.60& 13.98$\pm$1.47& 0.12$\pm$0.00&
-0.10$\pm$0.05& 14.56$\pm$0.12& 82.52$\pm$0.66\\
14& 0.64& $\gid$0.20& 54.35$\pm$4.23& 11.68$\pm$0.91& 1.57$\pm$0.09& $\gid$0.20& 41.58$\pm$0.87& 63.72$\pm$1.34& 0.38$\pm$0.03&
-0.04$\pm$0.12& 4.07$\pm$0.07& 24.60$\pm$0.44\\
15& 0.71& -1.01$\pm$0.19& 45.57$\pm$3.01& 17.18$\pm$1.13& 2.17$\pm$0.26& $\gid$0.20& 29.69$\pm$1.46& 21.56$\pm$1.06& 0.71$\pm$0.02&
0.02$\pm$0.04& 24.74$\pm$0.08& 61.26$\pm$0.19\\
16& 0.53& 0.14$\pm$0.09& 76.61$\pm$3.46& 19.84$\pm$0.90& 1.43$\pm$0.13& $\gid$0.20& 10.47$\pm$0.87& 20.11$\pm$1.66& 0.60$\pm$0.04&
-0.02$\pm$0.06& 12.93$\pm$0.17& 60.05$\pm$0.78\\
17& 0.59& 0.19$\pm$0.10& 88.51$\pm$3.46& 15.11$\pm$0.59& 1.23$\pm$0.37& -1.54$\pm$0.17& 7.03$\pm$0.41& 25.45$\pm$1.48& 0.12$\pm$0.00&
-0.45$\pm$0.07& 4.46$\pm$0.07& 59.44$\pm$0.91\\
18& 0.59& $\gid$0.20& 62.20$\pm$2.47& 8.86$\pm$0.35& $\lid$1.00& 0.02$\pm$0.03& 32.33$\pm$0.49& 51.33$\pm$0.78& 0.16$\pm$0.02& 0.14$\pm$0.12&
5.46$\pm$0.06& 39.80$\pm$0.44\\
19& 0.55& -1.24$\pm$1.02& 7.56$\pm$4.58& 1.85$\pm$1.12& 1.64$\pm$0.09& 0.19$\pm$0.06& 66.84$\pm$1.53& 42.87$\pm$0.98& 0.49$\pm$0.02& 0.01$\pm$0.05&
25.60$\pm$0.10& 55.28$\pm$0.21\\
20& 0.51& $\gid$0.20& 31.64$\pm$16.14& 3.47$\pm$1.77& 1.58$\pm$0.37& 0.17$\pm$0.12& 29.93$\pm$3.90& 23.23$\pm$3.03& 0.68$\pm$0.05&
-0.09$\pm$0.06& 38.44$\pm$0.67& 73.30$\pm$1.27\\
21& 0.61& 0.19$\pm$0.18& 71.42$\pm$2.10& 15.63$\pm$0.46& 0.84$\pm$0.06& 0.10$\pm$0.04& 27.41$\pm$0.44& 75.05$\pm$1.21& 0.27$\pm$0.15&
-0.01$\pm$0.54& 1.18$\pm$0.10& 9.31$\pm$0.77\\
22& 0.60& $\gid$0.20& 86.07$\pm$3.29& 29.72$\pm$1.14& 1.24$\pm$0.15& -0.20$\pm$0.10& 11.63$\pm$0.57& 46.23$\pm$2.26& 0.32$\pm$0.07&
0.19$\pm$0.29& 2.30$\pm$0.11& 24.05$\pm$1.14\\
23& 0.53& -0.52$\pm$0.24& 52.46$\pm$6.69& 14.75$\pm$1.88& 1.44$\pm$0.09& 0.10$\pm$-0.00& 35.19$\pm$1.99& 46.17$\pm$2.62& 0.51$\pm$0.05&
0.17$\pm$0.05& 12.35$\pm$0.24& 39.08$\pm$0.76\\
24& 0.60& $\gid$0.20& 75.12$\pm$2.54& 15.15$\pm$0.51& 2.53$\pm$0.95& -1.29$\pm$0.37& 7.78$\pm$0.68& 15.30$\pm$1.33& 0.54$\pm$0.03&
0.01$\pm$0.05& 17.10$\pm$0.20& 69.55$\pm$0.82\\
25& 0.61& 0.17$\pm$0.14& 70.58$\pm$2.69& 15.57$\pm$0.59& 0.92$\pm$0.07& 0.17$\pm$0.08& 20.91$\pm$0.87& 49.71$\pm$2.06& 0.59$\pm$0.08&
0.01$\pm$0.10& 8.51$\pm$0.36& 34.72$\pm$1.49\\
26& 0.77& 0.16$\pm$0.07& 91.24$\pm$1.32& 32.26$\pm$0.47& 1.23$\pm$0.13& -1.51$\pm$0.28& 3.88$\pm$0.19& 28.16$\pm$1.39& 0.42$\pm$0.02&
0.20$\pm$-0.00& 4.88$\pm$0.12& 39.58$\pm$0.94\\
27& 0.64& $\gid$0.20& 92.86$\pm$0.91& 37.35$\pm$0.37& $\lid$1.00& -0.03$\pm$0.18& 2.42$\pm$0.26& 11.23$\pm$1.23& 0.55$\pm$0.06&
-0.72$\pm$0.08& 4.72$\pm$0.08& 51.42$\pm$0.89\\
28& 0.59& -0.66$\pm$0.19& 53.84$\pm$4.23& 16.91$\pm$1.33& 1.53$\pm$0.11& $\gid$0.20& 31.22$\pm$1.36& 36.58$\pm$1.59& 0.56$\pm$0.04&
0.01$\pm$0.09& 14.95$\pm$0.11& 46.51$\pm$0.33\\
29& 0.54& -0.54$\pm$0.23& 58.11$\pm$9.55& 13.66$\pm$2.24& 1.61$\pm$0.35& $\gid$0.20& 25.81$\pm$2.87& 23.15$\pm$2.58& 0.32$\pm$0.01&
$\gid$0.20& 16.07$\pm$0.09& 63.19$\pm$0.37\\
30& 0.62& -1.03$\pm$0.13& 58.69$\pm$2.43& 19.72$\pm$0.82& 1.24$\pm$0.05& 0.03$\pm$0.06& 32.93$\pm$0.82& 44.27$\pm$1.10& 0.31$\pm$0.01&
$\gid$0.20& 8.39$\pm$0.07& 36.01$\pm$0.31\\
31& 0.66& -0.47$\pm$0.14& 70.69$\pm$4.39& 28.42$\pm$1.76& 1.59$\pm$0.21& 0.14$\pm$0.08& 24.87$\pm$1.34& 41.16$\pm$2.22& 0.32$\pm$0.02&
$\gid$0.20& 4.44$\pm$0.07& 30.42$\pm$0.48\\
32& 0.86& -0.02$\pm$0.13& 85.35$\pm$1.81& 30.80$\pm$0.65& 2.61$\pm$0.66& $\lid$-1.68& 6.44$\pm$0.57& 20.54$\pm$1.81& 0.57$\pm$0.07&
0.00$\pm$0.13& 8.20$\pm$0.20& 48.65$\pm$1.18\\
33& 0.65& 0.19$\pm$0.17& 77.35$\pm$4.17& 18.26$\pm$0.99& 1.03$\pm$0.21& -0.22$\pm$0.13& 14.95$\pm$1.24& 44.67$\pm$3.70& 0.52$\pm$0.08&
0.08$\pm$0.10& 7.70$\pm$0.57& 37.08$\pm$2.73\\
34& 0.57& $\gid$0.20& 86.94$\pm$1.63& 29.53$\pm$0.55& 0.93$\pm$0.08& -0.05$\pm$0.13& 10.57$\pm$0.35& 45.15$\pm$1.51& 0.32$\pm$0.04&
0.20$\pm$0.10& 2.48$\pm$0.10& 25.32$\pm$0.97\\
35& 0.61& 0.09$\pm$0.14& 78.63$\pm$3.07& 19.70$\pm$0.77& 0.97$\pm$0.10& 0.02$\pm$0.11& 15.92$\pm$0.70& 42.08$\pm$1.86& 0.32$\pm$0.02&
0.20$\pm$0.10& 5.45$\pm$0.16& 38.22$\pm$1.11\\
\hline
\end{tabular*}
\end{table*}

\begin{figure*}
\centering
 \includegraphics[scale=0.6]{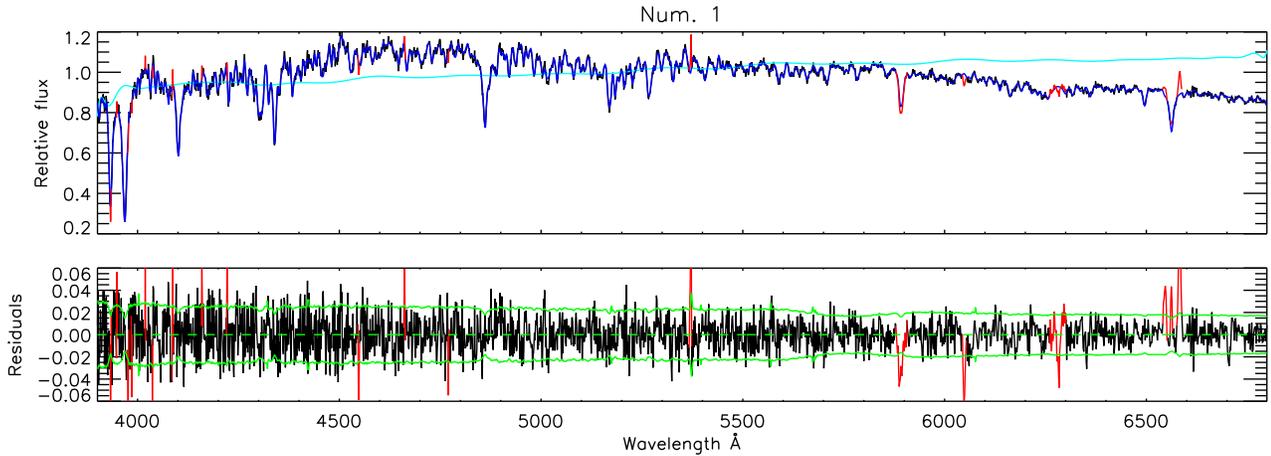}
 \caption[]{Best fit with the PE model and the residual spectrum for Galaxy number 1.
The top panel shows the spectrum in black and the best fit in blue (both are almost superimposed and the
black line can be seen only when zooming on the figure), the cyan line is the multiplicative polynomial
to absorb the effects of an unprecise flux calibration and of the Galactic extinction. The importance and details about the multiplicative polynomial 
are discussed in \citet{b18}. The red regions are rejected from the fit (rejection of flagged telluric
lines and automatic rejection of outliers). The bottom panel is the residuals from the best fit.
The continuous green lines mark the 1-$\sigma$ deviation, and the dashed green line is the zero-axis.}
\end{figure*}


\begin{figure*}
\centering
 \includegraphics[scale=0.8]{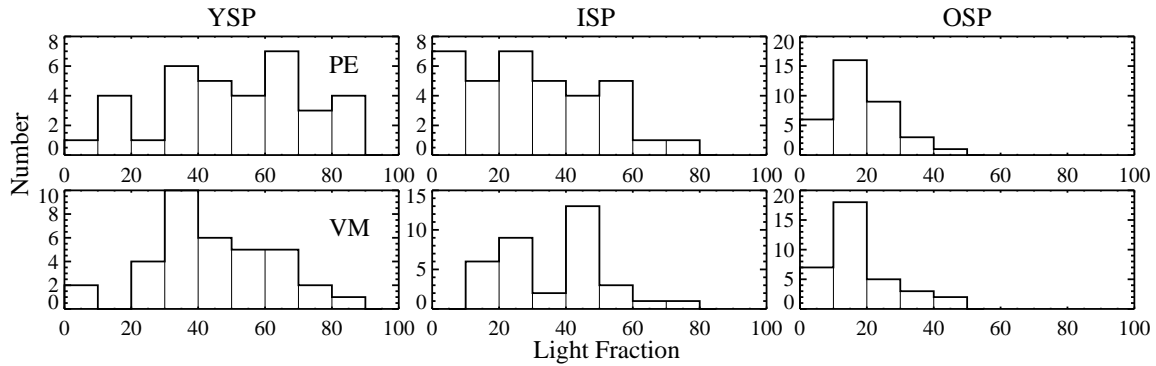}
 \caption[]{Distributions of the 3 populations with respect to light fraction (LF). The top panel shows LF distributions of YSP (left), ISP (middle), and OSP (right) from 3-population fit with the PE model. The bottom panel shows the same but with the VM model. For the top panel, the mean LFs of all YSPs, ISPs and OSPs are 51$\%$, 30$\%$ and 19$\%$. 
For the bottom panel, the mean LFs are respectively 46$\%$, 35$\%$ and 19$\%$. 
These values are tabulated in Table 7 for a direct comparison. 
 }
\end{figure*}

\begin{figure*}
\centering
 \includegraphics[scale=0.8]{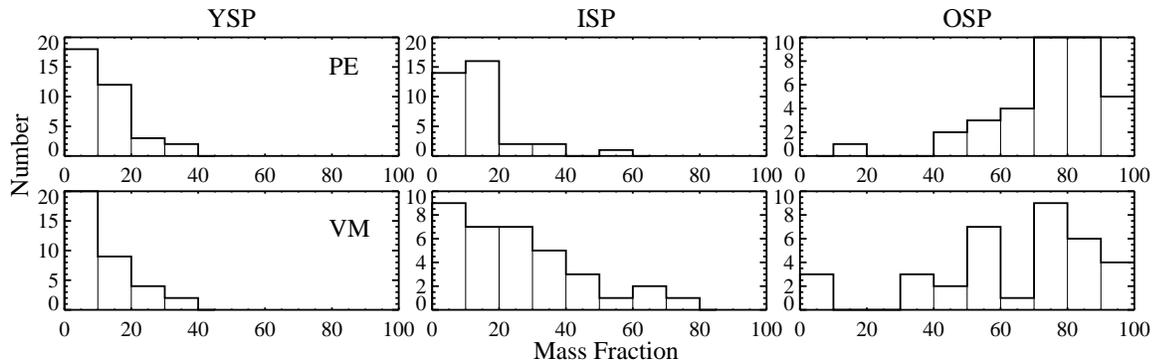}
 \caption[]{Distributions of the 3 populations with respect to mass fraction (MF). 
The orientation is the same as that in Figure 9.
For the top panel, the mean LF of all YSPs, ISPs and OSPs are 11$\%$, 14$\%$ and 75$\%$. 
For the bottom panel, the mean LFs are respectively 12$\%$, 25$\%$, and 63$\%$. 
These values are tabulated in Table 7.
 }
\end{figure*}

\section{Reliability of the solutions}
It is legitimate to question the reliability of the solutions because of degeneracies, so in this section we are demonstrating the reliability of our solutions in terms of the robustness of ULySS techqiue, the validity of our solutions, the sensitivity to the population model and the comparison with another work.

\subsection{Robustness of synthesis technique}  
ULySS is a package to fit spectroscopic observations against a linear combination of non-linear model components convolved with a parametric line-of-sight velocity dispersion (LSF function in $\S$ 3.1). It fits the full spectrum with population models to derive the physical properties of stellar populations. This method appears as an optimized alternative to methods based on spectrophotometric indices or spectral energy distribution (SED), as it ensures a high utilization ratio of all the useful information contained in the signal by fitting all the pixels. The algorithm has been optimized in the mathematical details to improve the precision, robustness and performance, which finds out the optimal values of unknown parameters and the errors on them. The optimal age and metallicity for each component result from the MPFIT function performed by the Levenberg-Marquardt least-squares minimization, and the optimal weight and luminosity weight of each component and the errors on them are determined by a multiple linear regression fit between the components and the observations (errors on the optimal age, [Fe/H], mass fration and light fraction of the components are all given in Table 5 and 6). It is convinced that results from ULySS method are consistent with those from only fitting Lick indices, but are more precise\citep{b18}. 

It is an old problem that there are multiple solutions existing in stellar population synthesis. The multiplicity is caused by a combination of three factors: algebraic degeneracy, intrinsic degeneracies of stellar populations, and the measurement uncertainties. The algebraic degeneracy arises from number of unknowns larger than number of observables. Unlike methods in which only a few spectral indices are used for synthesis, ULySS synthesize the full spectrum, so the algebraic degeneracy is not a problem, as the number of the points used in the observational spectrum  far exceed the number of unkown parameters. Similarly, the degeneracies associated with different stellar populations should be relieved by fitting the full spectrum\citep{b37}. Furthermore, ULySS determine all the free parameters in a single fit so as to effectively deal with the existing degeneracies between them instead of estimating the parameters in different steps. More explicit explanation on this idea of handling degeneracies can be read in Section 2 in \citet{b11}.

To check the reliability of ULySS, \citet{b18} have analysed the spectra of Galactic clusters whose populations are already known from colour-magnitude diagrams (CMD)determinations, and the two results are well consistent. So far, ULySS has been presented in several places to study stellar populations and its reliability assessed on the basis of various simulations \citep{b18,b35,b36}. 

To carefully verify the reliability and robustness of ULySS by ourselves, we analyse the ``stellar populations" of a `fake galaxy' made from known SSPs. Such a test has also been adopted by STARLIGHT group to check the robustness of STARLIGHT \citep{b38}. In our test, we firstly extract three spectra with definite ages and metallicities from the PE model. Spectrum 1 represents an old stellar population with age=10 Gyr and [Fe/H]=-1.0 dex. Spectrum 2 represents an intermediate-age stellar population with age=2500 Myr and [Fe/H]=-0.5 dex. Spectrum 3 represents a young stellar population with age=500 Myr and [Fe/H]=0.2 dex. For a convenient use, we represent Spetrum 1, 2 and 3 as Old, Inter and Young, respectively. Secondly, we generate a composite `fake galaxy' spectrum by the linear combination of 0.1$\times$young, 0.3$\times$inter and 0.6$\times$old (composite spectrum=0.1$\times$young+0.3$\times$inter+0.6$\times$old). Finally, we use ULySS to synthesize the `fake galaxy' and derive its components. The results have a good agreement with the known components (See Table 8).

\begin{table}
  \caption{Mean light fraction, mass fraction and age for the three populations of the 35 brightest E+A galaxies, illustrated respectively from the PE and VM model. }
\setlength{\tabcolsep}{1.0pt}
\renewcommand{\arraystretch}{1.1}
  \begin{tabular*}{84mm}{@{}l|ccc|ccc|ccr}
\hline
 Model &\multicolumn{3}{|c}{Mean light fraction} &\multicolumn{3}{|c}{Mean mass fraction} &\multicolumn{3}{|c}{Mean age (Gyr)}\\
       &YSP &ISP &OSP   &YSP &ISP &OSP  &YSP &ISP &OSP     \\
 
 \hline
PE  &51$\%$ &30$\%$ &19$\%$  &11$\%$ &14$\%$ &75$\%$  &0.54 &1.82 &12\\
VM  &46$\%$ &35$\%$ &19$\%$  &12$\%$ &25$\%$ &63$\%$  &0.46 &1.82 &12\\
\hline
\end{tabular*}
\end{table}
\begin{table}
  \caption{Comparison of the known and synthesized information for `fake galaxy'.
            }
  \begin{tabular*}{85mm}{@{}ccccc}
\hline
 \multicolumn{5}{c}{The known 'fake galaxy'}\\
\hline
SSP & age(Gyr) &[Fe/H](dex) &Weight &    \\
 Old &10.00 &-1.00 &0.60 &\\
Inter&2.50 &-0.50 &0.30&\\
Young&0.50 &0.20 &0.10&\\
 \hline\hline
\multicolumn{5}{c}{The synthesis results by ULySS}\\
\hline
SSP & age(Gyr) &[Fe/H](dex) &Weight  &light fraction   \\
Old &10.00$\pm$1.77 &-1.00$\pm$0.12 &0.60$\pm$0.03 &28.64$\pm$1.30\\
Inter&2.50$\pm$0.32 &-0.50$\pm$0.11 &0.30$\pm$0.01 &34.33$\pm$1.40\\
Young&0.50$\pm$0.02 &0.20$\pm$0.03  &0.09$\pm$0.00 &37.03$\pm$0.13\\

\hline
\end{tabular*}
\end{table}

\begin{figure}
 \includegraphics[scale=0.6]{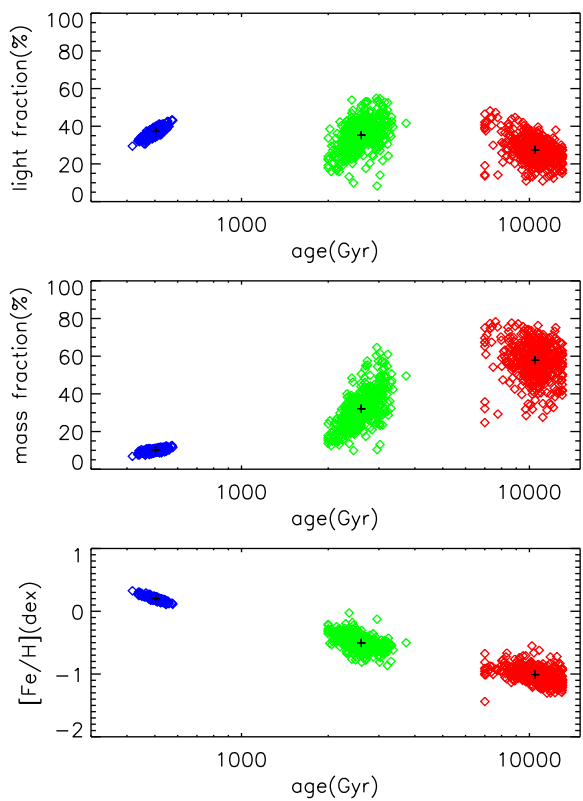}
 \caption[]{500 Monte-Carlo simulations for the `fake galaxy'. From top to bottom, in each picture, Young components are represented by blue small squares (the left gathering area). Intermediate-age components are represented by green small squares (the middle gathering area). Old components are represented by red small squares (the right gathering area).}
\end{figure}

In addition, we make 500 Monte-Carlo simulations which could visualize the degeneracies for this solution. In every simulation, we add a random Gaussian noise to the `fake galaxy' and then synthesize it by ULySS. From Monte-Carlo simulations presented in Figure 6, the YSPs (the left blue gathering area) concentrate at its true place (age=500Myr and [Fe/H]=0.2dex), compactly. However, the ISPs (the middle green gathering area) and OSPs (the right red gathering area) scatter a little, but anyway they can be divided into two different components. The gathering centers (plus sign) of simulations are consistent with our fit results, which are the closest to the true value (see Table 8 for comparison).  Thus we believe that ULySS can at least give the most likely solutions. This is enough for us, because we just would like to find out the dominant population and some quanlitative conculsions.

\subsection{Validity of our solutions }

To check the reliability of our solutions, we deal the 35 E+A sample galaxies as a whole, and combine their spectra to be one composite spectrum by using IRAF task-SCOMBINE. This combined spectrum could be a good representative of E+A galaxies. 

For this combined spectrum, we do the same 3-population fit with the PE model as made in $\S$3.3. The results are listed in Table 5 (the last line). Then we make 500 Monte-Carlo simulations for this combined spectrum to verify our solutions through visualizing the degeneracies (see Figure 7). During each simulation, a different random Gaussian noise is added to the galaxy and then the 3-population fit with the PE model is made. In Figure 7, the 500 results of each component scatter, but they all respectively gather towards a center (black plus sign) which is consistent with our fit results. The three components (young, intermediate and old) represented by different colours (blue, green and red) and gathered respectively in the left, middle, and right area in every picture can be clearly seperated . Although scatters in each component exists and then the age of each component can move around within the gathering area filled with its colour, in the top picture (relation of age and light fraction), the distribution of YSPs (the left blue region) are statistically at the top, ISPs (the middle green region) at the middle, and OSPs (the right red region) at the bottom, along the vertical axis (light fraction). This means in a statistical sense, small changes in age do not affect the dominance relationship in light among YSP, ISP and OSP. Similarly, the middle picture convinces that the OSPs statistically contribute the most to the mass, and the YSPs contribute the least. These solutions from simulations are consistent with our previous solutions for the 35 brightest E+A galaxies, that is, to light, YSPs are dominance, then ISPs and finally the OSPs; to mass, OSPs are dominance, then ISPs and finally the OSPs.

For a more clear comparison, we list the fit result (F) copied from the last line of Table 5 and the mean result from our 500 simulations (S) for this combined spectrum in Table 9. From Table 9, we can see that for the combined spectrum, the mean results from the simulations are generally consistent with the fit results listed in Table 5 (the last line). For this combined spectrum, no matter which result is considered, F or S, YSPs contribute about 60$\%$, ISPs about 30$\%$ and OSPs about the rest 10$\%$ of the light. However, OSPs contribute about 50$\%$, ISPs about 30$\%$ and YSPs about the rest 20$\%$ of the mass. This is completely corresponding with our previous statistical solutions. Such consistency can underpin the correctness of our solutions, since this analysed combined spectrum represent all the E+A sample galaxies as a whole.

  



\begin{table*}
 \caption{Comparison between the fit results (F) and the 500 simulations results (S) for the combined spectrum. The first line shows the optimal results copied from Table 5, and the second line shows the mean results (the position of the plus sign in Figure 7) from 500 simulations.
          }
\setlength{\tabcolsep}{1.0pt}
\renewcommand{\arraystretch}{1.1}
\begin{tabular*}{170mm}{@{\extracolsep{\fill}}c|cccc|cccc|cccc}
\hline
Result& \multicolumn{4}{c|}{OSP (fixed to 12Gyr)} & \multicolumn{4}{c|}{ISP}         & \multicolumn{4}{c}{YSP}  \\
\cline{2-13}
            &Age   & [Fe/H]&MF     &LF                    & Age & [Fe/H]&MF    & LF     & Age & [Fe/H] &MF    & LF\\
            &(Gyr) &(dex)  &($\%$) &($\%$)                &(Gyr)&(dex)  &($\%$)&($\%$)  &(Gyr)&(dex)   &($\%$)&($\%$)\\
  
\hline\hline
F  &12.00 &-0.20 &57.95 &13.05 &1.35 &0.39 &22.44 &26.80 &0.72 &-0.22 &19.61 &60.14\\
S  &12.00 &-0.26 &49.67 &11.28 &1.56 &0.30 &32.73 &34.95 &0.67 &-0.18 &17.60 &53.77\\

\hline
\end{tabular*}
\end{table*}

\begin{figure}
 \includegraphics[scale=0.6]{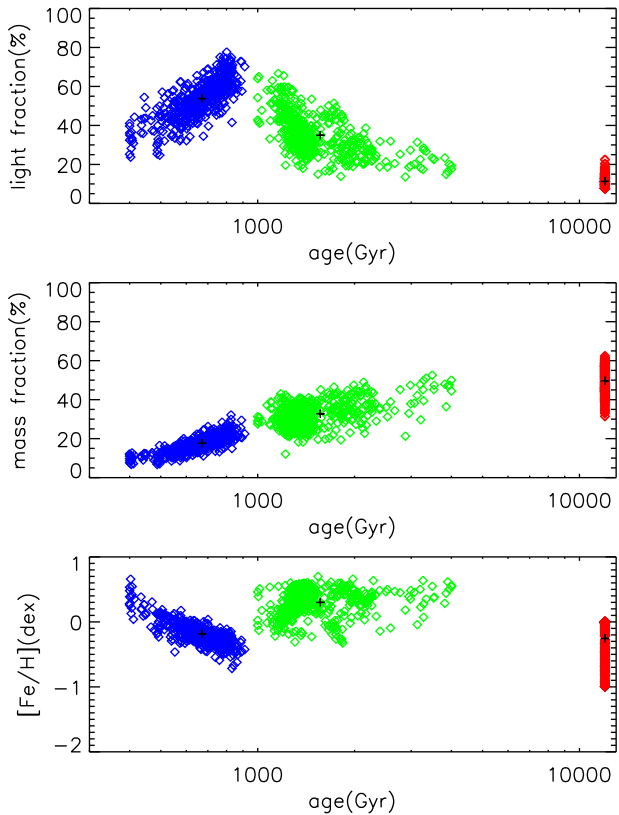}
 \caption[]{500 Monte-Carlo simulations for the combined spectrum. From top to bottom, in each picture, Young components are represented by blue small squares (the left gathering area). Intermediate-age components are represented by green small squares (the middle gathering area). Old components are represented by red small squares (the right gathering area).}
\end{figure}

Another way to check the reliability of our solutions is to fix the fitting technique but change the population models to fit our sample. In our work, the PE and VM model are two different and independent population models. We have fitted all of our sample using both of the models. The agreement of the results from the two models is satisfactory (see details in $\S$ 3.2 and $\S$ 3.3).

Comparing our solutions with those from other work is also a good way to verify the reliability of our solutions. Our work has shown in Table 7 that the optical light of E+As are dominated by YSPs ($< $1Gyr), the average light fraction of which is the most of all (51$\%$ illustrated with the PE model and 46$\%$ illustrated with the VM model). The average light fraction is less in the ISPs (30$\%$ with the PE model and 35$\%$ with the VM model), and the least in OSPs (19$\%$ with the PE model and 19$\%$ with the VM model). Our solutions are consistent with those of \citet{b14} for the stellar population study of the 2dF E+A sample galaxies. Their work has revealed that the composite model spectra that best fit the E+A spectra are dominated by YSP templates ($< $1Gyr). The agreement between the solutions from our work and other work on E+As further underpins the reliability of our solutions.

All this series of tests and comparisons could verify that our solutions are robust.
 
\section{Reconstruction of the Star formation histories of the sample$-$SFR}
The fundamental information that characterizes CSPs is its SFH, that is the evolution with time of the amount (i.e. total mass) of stars formed (SFR). The traditional and simplest way to probe the galaxy evolution is to measure the observed SFR in galaxies. 

ULySS does not imediately give the SFR after fitting, but it gives the optimal weights of the different SSP components. To derive the SFR we have to assume two points. Firstly, the individual SSPs  approximate the SFH over a period of time.  Secondly, we regard the SFR as a constant in each time box. If only these two assumptions are taken into account, we can then calculate the SFR.

First of all, we have to choose the limits of these time boxes. We set limits at medium location between the logarithm ages of the components, and derive the outer limits so that the age of the extreme components sit in the middle of their boxes. This approach is the smoothing strategy, which associates a duration to each burst and then smooths the instantaneous burst in its assuming duration time box by convolving with a Gaussian function. The choices of the time box above is nearly the centers of the Gaussian function. We consider the 12Gyr in the past as the beginning of the star formation. Such choices for time boxes have already been adopted in \citet{b35} and \citet{b39} to get generally smoothing star formation histories. 

Then, the smoothing SFR in each time box is calculated out as the weight of the correlated stellar component divided by the length of the time box.  Thus, we obtain the smoothing SFRs for all of the 35 E+As. In Figure 8, we present the SFRs for each of the sample as a function of evolutionary time. For each of the sample, the SFR illustrated from the fit with the PE model is represented by the solid line while SFR illustrated from the VM model is represented by the dashed line. When we see Figure 8, all the assumptions above have to be taken into account. This figure at least allows a direct comparison of the histories reconstructed with the two different models. 

\begin{figure*}
\centering
\includegraphics[scale=1.3]{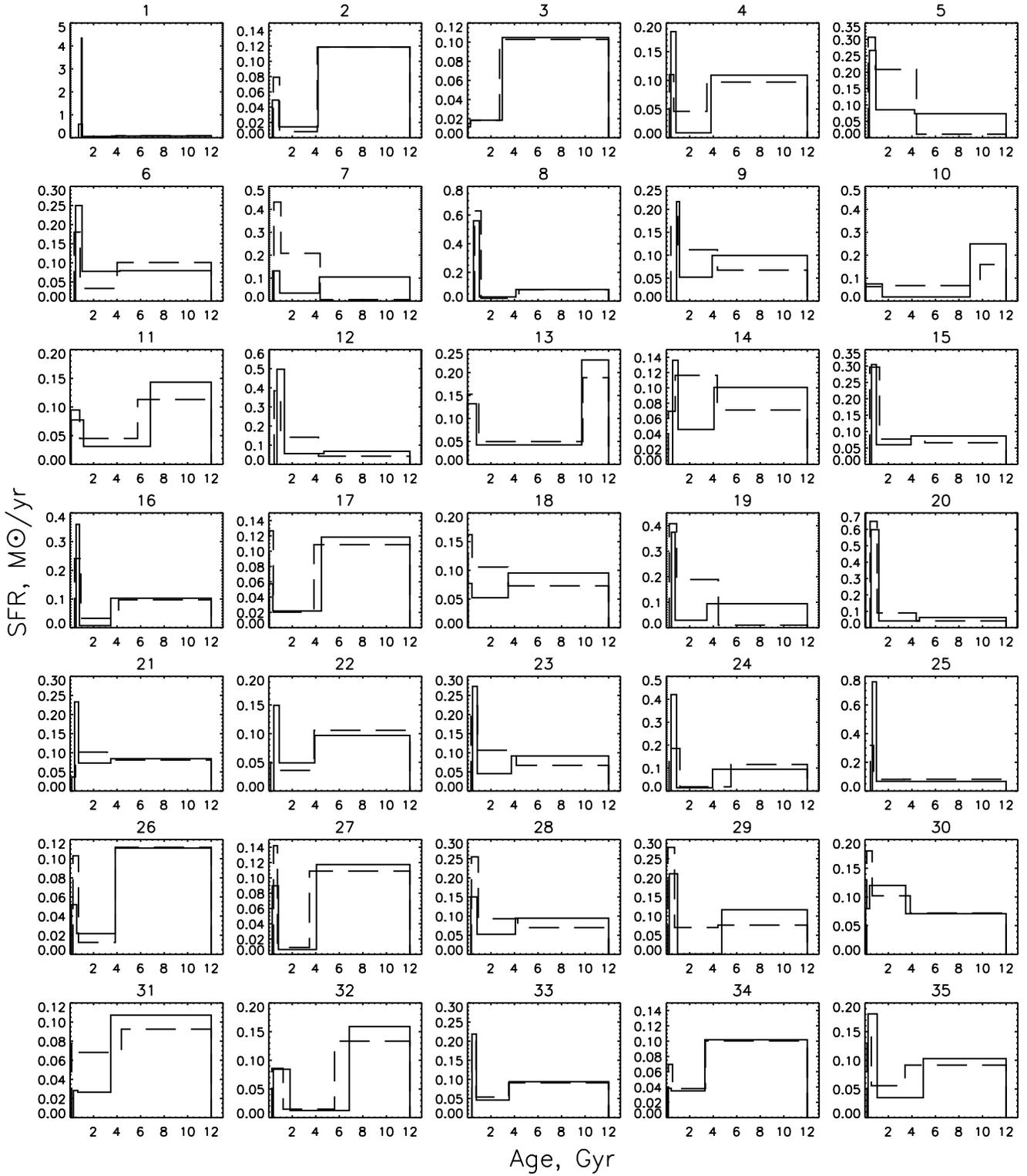}
\caption[]{SFR illustrated from the fit with the PE model (in solid line) and the VM model (in dashed line) for each of the sample.}
\end{figure*}

In Figure 8, the SFR for each of the sample have good agreement between the PE model and the VM model. Almost all of the sample galaxies have the relatively low SFR in the intermediate evolutionary stage, and less than 0.1 solar mass of stars is formed stably every year in this stage. For 11 E+A galaxies (Galaxy number 2, 3, 10, 11, 13, 17, 26, 27, 31, 32 and 34), the early stage is important for their formation, as they have high SFRs, which are greater than 0.1 solar mass every year in the early period. For other 11 E+A galaxies (Galaxy number 1, 4, 5, 6, 8, 9, 12, 15, 16, 19 and 20), the SFRs during the late evolutionary stage are quite high, and at least 0.15 solar mass of stars can be formed in this stage. Such a phenomenon could be due to the starburst in the E+A galaxy which takes place recently and is truncated in a short time. For these 11 E+As, it is probable that the timescale for the lateset formation phase is much shorter than assumed and the instantaneous SFR of this phase is much higher. These are just our explanation and we expect to explore more in our future work.

\section[]{Evolutionary scenario}
It is commonly accepted that E+A galaxies are the transformation phase from late-type galaxies to early-type galaxies,
and they are likely to evolve into E/S0 galaxies, which are the typical early-type galaxies \citep{b13,b12,b29}. However, the true nature and the validity of the link between E+As and E/S0s remain uncertain, while some authors make their works
to increase the reliability of this link. \citet{b12} have used high spatial resolution images to probe the detailed
morphologies of the local E+A galaxies, and they find that the properties of morphologies, color profiles, scaling 
relations, and young star clusters of E+As are either consistent with those of E/S0s, or indicate that E+A galaxies are 
caught in the act of transforming from late-type to early-type galaxies. \citet{b14} also prove the evolutionary link
from the evidence of the morphology. Their E+A galaxies are consistent with being early-type systems based both on radial
surface brightness profiles and visual morphological classification. Besides, the ubiquitous rotation found in their E+A
galaxies provides that E+A galaxies are completely consistent with the `fast rotator' population of early-type galaxies.
This gives another evidence to prove the evolutionary link between the E+As and E/S0 (early-type galaxies).

In this work, we expect to give another evidence from ages of the stellar populations to support that evolutionary link between E+As and early-type galaxies. If there is sure to exist the evolutionary link from E+As to early-types (E/S0s), we would expect the early-types have, in general, older old stellar population ages than E+As. 

To complete it, we plan to fit our E+As and 34 SDSS early-type galaxies (ellipticals) randomly selected from \citet{b27} with 2-population models by ULySS since we hope to seperate their components into two groups in age (one is the old age group, and the other is the young age group), and then we compare the equivalent age of the old components from our E+A sample and the selected ellipticals. It is expected that the equivalent age of old components in ellipticals is older than that in E+As. If this hypothesis is true, an evidence in stellar population ages can be obtained to support the evolutionary link.
 While we are fitting the E+A sample, we choose SSPs aged between 10Myr and 2Gyr as young components, and SSPs aged between 2Gyr and 16Gyr as old components, and we fit each of the E+A sample with these two components by using ULySS with the PE model. The age limits for two components are fine for all of E+A galaxies. We obtain the equivalent ages of old component and young component for each of the sample, and the validity of the solution is under the robustness of ULySS ($\S$ 4.1) and is also tested by the same Monte-Carlo simulations as described in $\S$ 4.2. For elliptical galaxies, if we use the same age limits for components, the age of young component would always reach the upper limit (2Gyr) after fitting, which reveals an older than 2Gyr age is required for young component of ellipticals. So we choose SSPs aged between 10Myr and 5Gyr as young components and SSPs aged between 5Gyr and 16Gyr as old components to fit ellipticals. These new limits are fine for all the ellipticals.   

We obtain the equivalent age of the old component ($t_{\rmn{O}}$) for each of the E+A galaxies and elliptical galaxies. Our results show that $t_{\rmn{O}}$ of 30 E+As (30/35=86$\%$) distribute lower than 10Gyr, and the average age of the old stellar populationsof E+As is concentrating on 7.0Gyr. However, for ellipticals 30 of them (30/34=88$\%$) have $t_{\rmn{O}}$ larger than 9Gyr, and the average age of the old stellar populations of the ellipticals is 11.5Gyr, which is much older than that of E+As (see distributions in Figure 9). 

Therefore, our workis consistent with the proposed evolutionary link from E+As to early-types (E/S0s) in the aspect of stellar population age. 


\begin{figure}
 \includegraphics[scale=0.4]{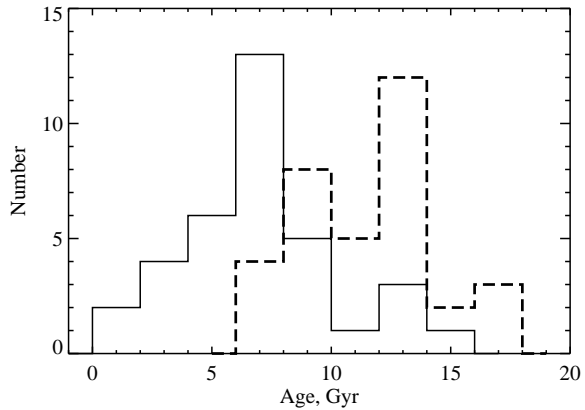}
 \caption[]{Distribution histogram of old stellar population ages for E+As and ellipticals from 2-population fit with the PE model. We use bin=1Gyr for this histogram. The solid line shows the age distribution of old components for E+As,
and the dashed line shows for ellipticals. The mean old component age for E+As is 7.0Gyr, and for ellipticals is 11.5Gyr. }
\end{figure}

\section[]{Summary and conclusions}
In this paper we have presented a detailed stellar population synthesis work on our 35 brightest E+A galaxies selected from Goto's E+A galaxies catalogue from SDSS DR5 \citep{b41}. These E+As have petrosian magnitudes in r-band less than 16 mag and spectroscopic S/N ratio in the continuum of the r-band wavelength range greater than 30. The redshifts of them are all within z = 0.14. Considering our selection criteria (brightest), the selected 35 E+A sample are good representatives of the full 564 E+As from Goto's catalogue in terms of the distributions of redshift, absolute magnitude, diameter, velocity dispersion, SSP-equivalent age and metallicity. 

We have fitted the 35 E+A sample with a SSP model and CSPs model respectively from the PE and the VM model. The SSP-equivalent age, metallicity and velocity dispersion of each of the sample can be obtained by a SSP fit, which has a mean SSP-equivalent age of 0.87Gyr with the PE model and 1.14Gyr with the VM model, a mean [Fe/H] of -0.04 dex and -0.38 dex, and a mean velocity dispersion of 159 km/s and 141 km/s (see Figure 2). The results from the two different models are generally consistant. 

For the CSP fit, we use the combination of YSP + ISP + OSP and get the optimal parameters of YSP, ISP and OSP for each of the sample. We have detected the 12Gyr OSPs with insignificant light fraction for all of the sample galaxies. However, the YSP components ($<$ 1Gyr) dominate the E+As in light, but contribute very little to the mass of E+A galaxies. Our conclusions are that for the E+A sample, YSPs are statistically the dominance to the light, then ISPs and finally OSPs. However, to mass, OSPs are statistically the dominance, then ISPs and finally YSPs.

 We verified the reliability and robustness of our solutions from 4 aspects: the robustness of the technique we used, the validity of our solutions, the consistency of the results from two independent population models and the agreement with other studies. 

We have reconstructed the SFH for each of the E+A sample by giving their smoothing SFR. For 11 E+As, the early stage is the significant period for star formation. However, for another 11 E+As, their SFRs in the late evolutionary stage are very high, which possibly arises from the extremely sudden truncation of the recent starburst.

At last, we additionally compared the stellar populations of the E+As with those of the early-type galaxies. We fitted our 35 E+As and the 34 early-types randomly selected from SDSS with 2-population model (OSP + YSP) from the PE model. The results show that the ellipticals have in general older OSP ages than E+As. This gives at least another evidence in the aspect of stellar population age to support the proposed evolutionary senario of E+As evolving into early-types (E/S0s).

\section*{Acknowledgments} 
We would like to thank the referee for very useful comments and suggestions and thank Xianmin Meng, Fengfei Wang, Xiaoyan Chen of National Astronomical Observatories, Chinese Academy of Sciences (NAOC) and Dr. Zhongmu Li of Dali College for their kindly help. This work is supported by the Natural Science Foundation of China (NSFC) under Nos. 10973021, Nos. 10933001 grants, the National Basic Research Program of China (973 Program No. 2007CB815404 and No. 2006AA01A120 (863 project) and the Young Researcher Grant of National Astronomical Observatories, Chinese Academy of Sciences.

\bsp

\label{lastpage}

\end{document}